\documentclass[aps,showpacs,amsmath,amssymbd,dvips,showkeys,preprintnumbers]{revtex4}

\usepackage{graphicx}
\usepackage{color}

\def \be  {\begin{equation}}
\def \ee  {\end{equation}}
\def \ee  {\end{equation}}
\def \bea {\begin{eqnarray}}
\def \eea {\end{eqnarray}}

\newcommand{\nn}{\nonumber}

\begin{document}

\preprint{ECTP-2015-07}    
\preprint{WLCAPP-2015-07}
\vspace*{3mm}

\title{Friedmann inflation in Horava-Lifshitz gravity with a scalar field}

\author{Abdel Nasser Tawfik\footnote{http://atawfik.net/}}
\affiliation{Egyptian Center for Theoretical Physics (ECTP), Modern University for Technology and Information (MTI),
 11671 Cairo, Egypt}
\affiliation{World Laboratory for Cosmology And Particle Physics (WLCAPP), Cairo-Egypt}

\author{Abdel Magied Diab}
\affiliation{Egyptian Center for Theoretical Physics (ECTP), Modern University for Technology and Information (MTI),
 11671 Cairo, Egypt}
\affiliation{World Laboratory for Cosmology And Particle Physics (WLCAPP), Cairo-Egypt}

\author{Eiman Abou El Dahab}
\affiliation{Faculty of Computer Sciences, Modern University for Technology and Information (MTI),
 11671 Cairo, Egypt}
\affiliation{Egyptian Center for Theoretical Physics (ECTP), Modern University for Technology and Information (MTI), 11671 Cairo, Egypt}

\date{\today}

\begin{abstract}

We study Friedmann inflation in general  Horava-Lifshitz (HL) gravity with detailed and non-detailed but also without the projectability conditions. Accordingly, we derive the modifications in the Friedmann equations due to single scalar field potentials describing power-law and minimal-supersymmetrically extended inflation. By implementing four types of the equations-of-state charactering the cosmic background geometry, the dependence of the tensorial and spectral density fluctuations and their ratio on the inflation field is determined. The latter characterizes the time evolution of the inflation field relative to the Hubble parameter. Furthermore, the ratio of tensorial-to-spectral density fluctuations is calculated in dependence on the spectral index.  The resulting slow-roll parameters apparently differ from the ones deduced from the standard General Relativity (Friedmann gravity). We also observe that the tensorial-to-spectral density fluctuations continuously decrease when moving from non-detailed HL gravity, to Friedmann gravity, to HL gravity without the projectibility, and to detailed HL gravity. This regular patter is valid for three types of cosmic equations-of-state and different inflation potential models. The results fit well with the recent PLANCK observations. 

\end{abstract}


\maketitle
\tableofcontents



\section{Introduction}
\label{sec:intr}

It is widely believed that the Universe in its very early stages went through a rapid exponential expansion \cite{infl}. At the beginning of this inflation epoch, scales inside the Hubble radius expanded, exponentially, as well \cite{inflnt1,inflnt2}.  At a later time,  due to large-scale cosmological perturbations with a scale-invariant spectrum, they reentered the Hubble radius.
This scenario on the inflation era \cite{infl1,infl2,Linde:1982} disfavors exotic relics such as primordial magnetic monopoles \cite{monop}, which otherwise would alter the present Universe.  Also, it was assumed that the exponential expansion could eliminate the particle horizon and therefore helps in solving the horizon problem \cite{inflhp}. Furthermore, the inflation scenario gives a solid explanation for flatness and homogeneity of our Universe and evolves quantum fluctuations in classical curvature perturbation, as well \cite{qfluct1,qfluct2}. It is conjectured that during the inflation era the perturbations started out with a small amplitude and gradually grown. This leads to a structure formation and thus even would explain the large-scale structure of the Universe. While detailed mechanism responsible for inflation is not well known yet, a number of predictions have been confirmed by observations \cite{inflobs1,inflobs2} and the hypothetical field inflaton \cite{inflaton} is thought as a likely candidate.

According to the cosmological equations suggested by the non-relativistic renormalizable gravity \cite{Kiritsis:2009a, Calcagni:2009s}, the early Universe gets features that may give an alternative to the inflation. The resulting theory, known as Horava-Lifshitz (HL) gravity would be able to avoid the Big Bang singularity \cite{Kiritsis:2009a}. Quantum corrections to gravity are conjectured to replace this singularity with an exponentially expanding de Sitter phase \cite{Starobinsky}. Recently, the HL gravity is assumed to introduce possible modifications in the standard General Relativity (GR) gravity, especially in strong gravitational regimes \cite{lifshitz,Horava2009a}. This is accompanied by modifications in the gravity itself making it responsible for the accelerated expansion of the Universe. Therefore, one of the natural implications of HL gravity is on the very early Universe. Furthermore, various modified theories of gravity, which are successfully able to unify different inflationary approaches with the late-time Universe acceleration and recent cosmological observations, can be implemented \cite{nori,masud}. 

It was demonstrated that the early-time inflation of the FRW Universe might take place as a power-law accelerating evolution or similar to de Sitter expansion \cite{Sasaki:2010a}. The earlier shall have a future singularity called sudden or Big Rip. When taking into consideration covariant gravity with a scalar field by an extra higher derivative term, which preserves the ultraviolet behavior of the graviton propagator, the future singularity can be cured and the Einstein frame-action \cite{Abdalla:2005}, where $f(R) \sim f_0 R^\alpha$, with constant $f_0$ and $\alpha$ should be modified, accordingly. Furthermore, the presence of $R^2$-term in the consistent modified gravity removes the future singularity \cite{Abdalla:2005}. The emergence of finite-time future singularities has been studied in Ref. \cite{1006.3387}. It was found that such singularities can be cured by adding a higher-order spatial derivative term. Recently, the general formulas for the inflationary power spectra of scalar and tensor are driven in the presence of a scalar field \cite{1409.1984}.

Based HL gravity without an additional scalar degree-of-freedom, a simple scenario for the scale-invariant quantum-fluctuations was proposed \cite{0904.2190} in order to impose inflation. For this simple scenario, the detailed balance conditions were not necessary but the inflation scenario itself may be still or no longer needed, for instance, serious horizon problems remain unsolved, such as monopole and domain walls. They still require inflation with slow-roll conditions.

In HL gravity without the projectability condition, the inflation regime was studied \cite{1208.2491}. But, opposite to the proposal of Ref. \cite{0904.2190}, the linear scalar perturbations equations of the FLRW universe are derived for a single scalar field. A master equation of the perturbations has been specified for a particular gauge. The power spectrum and spectrum index of the comoving curvature perturbations have been determined. It was noticed that the perturbations remain scale-invariance, and the HL gravity without the projectability condition is consistent with all current cosmological observations. This another solid support for adding scalar field(s).

In framework of nonrelativistic HL gravity with the projectability condition and an arbitrary coupling constant ($\lambda$), the inflation was studied \cite{1201.4630}. Accordingly, the FLRW Universe without specifying the gauge is necessarily flat. But by adding a single scalar field, it was noticed that both metric and scalar field become strongly coupled and almost identically oscillating in sub-horizon regions. In super-horizon regions, the comoving curvature perturbation remains constant although the FLRW perturbations become adiabatic. Furthermore, the perturbations in the slow-roll parameters, for instance, both scalar and tensor are found scale-invariant. Concrete tuning the coupling coefficients makes the spectrum index of the tensor perturbation identical as the ones deduced from  GR. But the ratio of scalar to tensor spectra can be similar to that from GR and seems to depend on the spatial higher-order derivative terms.

Following this line, we apply two models for the cosmic inflation and HL gravity. It intends to find out consistency with the recent cosmological observations and the possible distinguishability from GR gravity. We shall compare between the various corrections  proposed to reduce the independently coupling constants of the HL gravity theory. In section \ref{sec:HLG}, a short reminder to Horava-Lifshitz gravity shall be outlined. The modified Friedmann equations due to HL gravity are studied under non-detailed, section \ref{sec:HLGndb}, and detailed balance conditions, section \ref{sec:HLGdb} and that without the projectability condition, section \ref{sec:wtproj}. The cosmic inflation models are implemented in section \ref{sec:cim}. Section \ref{FluctuationHLG} is devoted to study the fluctuations and the  slow-roll parameters in HL gravity. The results shall be confronted to the recent PLANCK observations in section \ref{sec:planck}. Section \ref{sec:conc} is devoted to final remarks and conclusions.

\section{Horava-Lifshitz Gravity}
\label{sec:HLG}

Starting from the dynamical variables, the lapse and shift vectors, $N$ and $N_i$, respectively, and canonical gravity, the full space-time metric reads 
\bea
ds^2 = - N^2 dt^2 + g_{ij}(dx^i + N^i dt)(dx^j + N^j dt),
\eea
where the indices are raised and lowered using the spatial metric $g_{ij}$. To explain quantum critical phenomena in condensed matter, Lifshitz proposed a scalar field theory \cite{lifshitz}. This has inspired Horava to propose a quantum gravity theory with an anisotropic scaling in ultraviolet (UV) \cite{Horava2009a}. 
Anisotropic UV scaling relations between Minkowskian space and time are thus the basic assumptions of this theory. In other words, the quantum gravity scaling at short distances exhibits a strong anisotropy between space and time
\begin{eqnarray}
t \longrightarrow  l^{z}\, t, \qquad & &
x^{i} \longrightarrow  l\, x^{i},
\end{eqnarray}
where $z$ is the dynamical critical exponent given in ultraviolet, and $l$ is a constant performing the scaling. As anisotropic scaling apparently implies a preferred time coordinate, $4$-dimensional Lorentzian metric can not be the only fundamental structure on Lorentzian (pseudo-Riemannian) manifold. In order to single out spatial (temporal) coordinates in a differentiable manifold, it was assumed that codimension foliation as a basic structure on this manifold  is resorted \cite{Horava2009a}. The foliation structure in turn fulfils local Galilean invariance but also implies impossible full-diffeomorphism invariance in GR.  This is recovered in some limits of the critical exponent $z$. Thus, GR is considered as an emerging in an infrared fixed point \cite{ir1}. To summarize, HL gravity is a projectable approach minimizing the number of independent couplings in potential and adopting an extra principle to build the potential, i.e. detailed balance condition.

The detailed balance condition is a technical treatment to further reduce the coupling constants, which would limit the prediction powers of the HL gravity theory. Furthermore, it was argued that the detailed balance HL gravity has to be broken in order to enable the theory to be compatible with the available observations \cite{Charm,Horava2011a}. A possible connection between detailed balance conditions and the entropic origin of gravity was discussed in Ref. \cite{Verlinde2011a}. Therefore, in constructing the potential, the detailed balance HL gravity can be implemented instead of effective field theory. 

The non-detailed balance condition was also proposed \cite{Horava2009a} in order to reduce the number of independently coupling constants \cite{const1a,const1b,const2}. These are likely when  general covariance is abandoned \cite{1208.2491}. This condition  assumes that the gravitational potential can be obtained from a superpotential defined on the three-spatial hypersurfaces $t=\mathtt{const}$, the Newtonian limit would not exist \cite{Reff6}, and a scalar field in UV could not be stable \cite{Reff7}. 

The projectability condition, which is usually taken into consideration with the balance conditions, assumes that the lapse function in the Arnowitt-Deser-Misner (ADM) decompositions is only time dependent. Recently, HL gravity without the projectability condition has been introduced \cite{1208.2491}.

In HL gravity, the general structure of the scalar field action is given by a consistent imposing of the corresponding symmetries \cite{Calcagni:2009s,Kiritsis:2009a}. However, one can in the construction of the action obtain the scalar field,  
\bea
\mathcal{S}^\phi &=& \int dt \, d^3x\, \sqrt{g} N \left[\frac{(3\lambda -1)\dot{\phi}^2}{4N^2}+ m_1 m_2 \phi \nabla^2 \phi - \frac{1}{2}m_2^2 \phi \nabla^4 \phi + \frac{1}{2}m_3^2 \phi \nabla^6 \phi - V(\phi) \right],
\eea
where $m_i$ are constants and $V(\phi)$ plays the role of an inflation potential.

\subsection{Non-detailed balance HL gravity}
\label{sec:HLGndb}

We shortly introduce the modified Friedmann equations in non-detailed balance HL gravity \cite{TDE}
\bea
H^2 &=& \frac{\kappa^2}{6(3 \lambda -1 )} \rho -\frac{\kappa^4 \mu^2 \Lambda ^2 _{\omega}}{16(3 \lambda -1)^2}+ 
 \frac{\kappa ^4 K \mu ^2 \Lambda _{\omega}}{8(3 \lambda -1 )^2 a^2}-\frac{\kappa ^4 K^2 \mu ^2 }{16(3 \lambda -1)^2 a^4}, \label{Freid1} \\ 
\dot{H} + \frac{3}{2} H^2 &=& -\frac{\kappa ^2}{4(3 \lambda -1 )} p -\frac{\kappa ^4 \mu ^2 \Lambda ^2 _{\omega}}{32(3 \lambda -1 )^2}+\frac{\kappa ^4 K \mu ^2 \Lambda _{\omega}}{16(3 \lambda -1 )^2 a^2},
\eea
where Hubble parameter $H=\dot{a}/a$, with $\dot{H}=\ddot{a}/a - H^2$. When we let $\xi=\kappa^2/(3 \lambda -1)$, Eq. (\ref{Freid1}) becomes 
\bea
\label{Fred2}
H^2 = \frac{\xi}{6} \rho - \frac{\xi ^2 \mu ^2 }{16} \left[ \Lambda ^2 _{\omega} - \frac{2 K \Lambda _{\omega}}{a^2} + \frac{K^2}{a^4} \right].
\eea

From first law of thermodynamics, the continuity equation can be given as
\bea 
\rho = \rho _o a^{-3 (1+\omega)},
\eea
which equivalently leads to $a \approx \rho^{-1/[3(1+ \omega)]}$. Then Eq. (\ref{Fred2}) can be written as
\bea
H^2 = \frac{\xi}{6} \rho - \frac{\xi ^2 \mu ^2 }{16} \left[ \Lambda ^2 _{\omega} - 2 K \Lambda _{\omega} \, \rho ^{\frac{2}{3(1+\omega)}} + K^2 \, \rho ^{\frac{4}{3(1+\omega)}} \right].
\eea

From classical field theory, the scalar field ($\phi$) in the total inflation energy is assumed to couple with the gravity  
\bea
\frac{1}{2} \left( \dot{\phi}^2 + (\nabla \phi)^2\, \right) + V(\phi). \label{infl_ener}
\eea
Motivation of introducing such scalar field was shortly detailed in introduction. 
Furthermore, the inflation dynamics can be described by various equations, for instance,
\begin{itemize}
\item Friedmann equation, which describes the contraction and expansion of the Universe, 
\bea
H^2 + \frac{\kappa\, c^2}{a^2}&=& \frac{8\, \pi\, G\, \rho + \Lambda\, c^2}{3}, 
\eea
especialy in homogeneous and isotropic Universe, and
\item Klein-Gordon equation, which is the simplest equation of motion for a spatially homogeneous scalar field
\bea
\ddot{\phi} + 3 \,H\, \dot{\phi} + \partial_\phi\, V(\phi) = 0, \label{KG}
\eea
where $\partial_\phi \equiv \partial/\partial \phi$. 
\end{itemize} 

The energy density sums up inflation and cosmological constant contributions, i.e. $\rho = \rho _{\phi} + \rho _{v}$. Then, the modified Friedmann equation becomes 
\bea
H^2 &=& \frac{\xi}{6}\left[ \frac{\dot{\phi}^2}{2} + V(\phi)+ \frac{ (\nabla \phi )^2}{2} +\rho _v \right] \nn \\
&-& \frac{\xi ^2 \mu ^2 }{16} \left\{ \Lambda _\omega ^2 -  2 K \Lambda _\omega \left(\frac{\dot{\phi}^2}{2} + V(\phi)+ \frac{ (\nabla \phi )^2}{2} +\rho _v \right)^{\frac{2}{3(1+\omega)}} \right.  \nn \\
&+& \left. K^2  \left( \frac{\dot{\phi}^2}{2} + V(\phi)+ \frac{ (\nabla \phi )^2}{2} +\rho _v \right)^{\frac{4}{3(1+\omega)}}\right\},
\eea
where $\rho_v=\Lambda\, \omega/8\, \pi \, G$ with $G$ being the gravitational constant. In rapidly expanding Universe and if the inflation field starts out sufficiently homogeneously, the inflation field becomes minimum, i.e. very slow \cite{Liddle:2003, Linde:2002}. This would be modelled by a sphere in a viscous medium, where both energy densities due to matter ($\rho_m$) and radiation ($\rho_r$) are neglected 
\bea
(\nabla \phi)^2 &\ll & V(\phi),  \label{eq:inql1} \\
\ddot{\phi} & \ll & 3\, H\, \dot{\phi}, \label{eq:inql2} \\
\dot{\phi}^2 & \ll &  V(\phi). \label{eq:inql3} 
\eea
The first inequality, Eq. (\ref{eq:inql1}), is obtained under the assumption of homogeneity and isotropy of Friedmann-Lemaitre-Robertson-Walker  (FLRW) Universe \cite{Liddle:2003, Linde:2002}, while the second inequality, Eq. (\ref{eq:inql2}), states that the scalar field changes very slowly so that the acceleration could be neglected \cite{Liddle:2003, Linde:2002}. The third inequality, Eq. (\ref{eq:inql3}), gives a principle condition for the Universe expansion. Accordingly, the kinetic energy is much less than the potential energy \cite{Liddle:2003, Linde:2002}. But it was observed that the Universe expansion apparently accelerates \cite{Linde:1982}.  Therefore, the modified Friedmann equation becomes 
\bea
H^2 &=& \frac{\xi}{6}\left[  V(\phi)+ \frac{\Lambda _\omega}{8\, \pi \, G} \right] - \frac{\xi ^2 \mu ^2}{16} \left[ \Lambda^2 _\omega + 
2 K \Lambda _\omega \left( V(\phi)+ \frac{\Lambda _\omega}{8\, \pi \, G} \right)^{\frac{2}{3(1+\omega)}} + 
K^2 \left( V(\phi)+ \frac{\Lambda _\omega}{8\, \pi \, G} \right)^{\frac{4}{3(1+\omega)}} \right], \hspace*{8mm} \label{eq:mdfdFRD}
\eea
which obviously depends on the equation of state:
\begin{itemize}
\item for dust approximation, i.e. $\omega=0$:\\
The corresponding modified Friedmann equation, Eq. (\ref{eq:mdfdFRD}), can be given as 
\bea
H^2 &=& \frac{\xi}{6}\left[  V(\phi)+ \frac{\Lambda _\omega}{8\, \pi \, G} \right] - \frac{\xi ^2 \mu ^2}{16} \left[ \Lambda ^2 _\omega - 
 2 K \Lambda _\omega \left( V(\phi)+ \frac{\Lambda _\omega}{8\, \pi \, G} \right)^{\frac{2}{3}} +K^2 \left( V(\phi)+ \frac{\Lambda _\omega}{8\, \pi \, G} \right)^{\frac{4}{3}} \right], \hspace*{8mm} 
\eea
\item for matter-dominated era, i.e. $\omega=1/3$ 
\bea
H^2 &=& \frac{\xi}{6}\left[  V(\phi)+ \frac{\Lambda _\omega}{8\, \pi \, G} \right] - \frac{\xi ^2 \mu ^2}{16} \left[ \Lambda ^2 _\omega  - 
 2 K \Lambda _\omega \left( V(\phi)+ \frac{\Lambda _\omega}{8\, \pi \, G} \right)^{\frac{1}{2}} +K^2 \left( V(\phi)+ \frac{\Lambda _\omega}{8\, \pi \, G} \right) \right], \hspace*{8mm}
\eea
\item for cold dark matter-dominated era, i.e. $\omega=-1/3$ 
\bea
H^2 &=& \frac{\xi}{6}\left[  V(\phi)+ \frac{\Lambda _\omega}{8\, \pi \, G} \right] - \frac{\xi ^2 \mu ^2}{16} \left[ \mu \Lambda ^2_\omega  - 
2 K \Lambda _\omega \left( V(\phi)+ \frac{\Lambda _\omega}{8\, \pi \, G} \right) +K^2 \left( V(\phi)+ \frac{\Lambda _\omega}{8\, \pi \, G} \right)^2 \right], \hspace*{8mm}
\eea
\item and for non-vanishing cosmological constant, i.e. $\omega=-1$ 
\bea
H^2 \longrightarrow \infty.
\eea
\end{itemize}

\subsection{Detailed balance HL gravity}
\label{sec:HLGdb}

The modified Friedmann equations due to detailed balance HL gravity is given as \cite{TDE}
\bea
H^2 &=& \frac{2}{(3\lambda - 1)} \left[ \frac{\Lambda _\omega}{2} + \frac{8 \pi G_N}{3} \rho - \kappa a^{-2} + \frac{\kappa ^2}{2 \Lambda _\omega} a^{-4} \right],
\eea
which reads the continuity equation, 
\bea
H^2 &=& \frac{2}{(3\lambda - 1)} \left[ \frac{\Lambda _\omega}{2} +  \frac{8 \pi G_N}{3} \left( V(\phi) + \frac{\Lambda _\omega}{8\, \pi \, G} \right) - \right. \nn \\
&& \left. \kappa  \left( V(\phi) + \frac{\Lambda _\omega}{8\, \pi \, G} \right)^{\frac{2}{3(1+\omega)}}+\frac{k^2}{2 \Lambda _\omega} \left( V(\phi) + \frac{\Lambda _\omega}{8\, \pi \, G} \right)^{\frac{4}{3(1+\omega)}} 
\right]. \hspace*{8mm} \label{eq:mdfdFRD2}
\eea
The various equations of state lead to
\begin{itemize}
\item for dust approximation, i.e. $\omega=0$:\\
The modified Friedmann equation, Eq. (\ref{eq:mdfdFRD2}), reads
\bea
H^2 &=& \frac{2}{(3\lambda - 1)} \left[ \frac{\Lambda _\omega}{2} +  \frac{8 \pi G_N}{3} \left( V(\phi) + \frac{\Lambda _\omega}{8\, \pi \, G} \right) - 
\kappa  \left( V(\phi) + \frac{\Lambda _\omega}{8\, \pi \, G} \right)^{\frac{2}{3}} + \frac{k^2}{2 \Lambda _\omega} \left( V(\phi) + \frac{\Lambda _\omega}{8\, \pi \, G} \right)^{\frac{4}{3}} 
\right], \hspace*{8mm}
\eea
\item for matter-dominated era, i.e. $\omega=1/3$ 
\bea
H^2 &=& \frac{2}{(3\lambda - 1)} \left[ \frac{\Lambda _\omega}{2} +  \frac{8 \pi G_N}{3} \left( V(\phi) + \frac{\Lambda _\omega}{8\, \pi \, G} \right) - 
\kappa  \left( V(\phi) + \frac{\Lambda _\omega}{8\, \pi \, G} \right)^{\frac{1}{2}}+\frac{k^2}{2 \Lambda _\omega} \left( V(\phi) + \frac{\Lambda _\omega}{8\, \pi \, G} \right)
\right], \hspace*{8mm}
\eea
\item for cold dark matter-dominated era, i.e. $\omega=-1/3$ 
\bea
H^2 &=& \frac{2}{(3\lambda - 1)} \left[ \frac{\Lambda _\omega}{2} +  \frac{8 \pi G_N}{3} \left( V(\phi) + \frac{\Lambda _\omega}{8\, \pi \, G} \right) - 
\kappa  \left(V(\phi) + \frac{\Lambda _\omega}{8\, \pi \, G} \right)+\frac{k^2}{2 \Lambda _\omega} \left(V(\phi) + \frac{\Lambda _\omega}{8\, \pi \, G} \right)^2 
\right], \hspace*{8mm}
\eea
\item and for non-vanishing cosmological constant, i.e. $\omega=-1$ 
\bea
H^2 \longrightarrow \infty.
\eea
\end{itemize}

\subsection{HL gravity without the projectability condition}
\label{sec:wtproj}

Both detailed and non-detailed corrections remarkably reduce the coupling constants. But essential problems in the HL gravity theory remain, such as strong coupling problem exists. Due to the existence of spin-0 graviton, Minkowski spacetime remains unstable, despite the de Sitter spacetime does. Spin-0 graviton causes difference of its speed from that of spin-2 graviton. Both are not related by any symmetry. Thus, any attempt to restore Lorentz symmetry at low energies is a big challenge, i.e. no mechanism ensuring that all species of matter and gravity have the same effective speed. Such problems have been discussed in Ref.  \cite{1208.2491}. HL gravity without the projectability condition was proposed to solve these problems. It is assumes that the detailed balance condition is softly broken, and the symmetry of the HL gravity theory is thus enlarged to include local U(1) symmetry \cite{modl1,modl2}. The latter eliminates spin-0 graviton. In flat FLRW space-time, the coupling constant $\Lambda_g$ vanishes \cite{1208.2491}, despite its radiative corrections expected from quantum mechanics. Then at vanishing cosmological constant, 
\bea
ds^2 &=& a^2(\eta) \left[ -d\eta^2 + \gamma_{ij}  dx^i dx^j\right],
\eea
where $\eta=\int (1/a) dt$ is the conformal time, and $\gamma_{ij} = \delta_{ij}/(1+ \kappa r^2/4)$. The modified Friedmann equation is
\bea
{^{\prime}H^2} = \frac{8\, \pi \tilde{G}\, a^2}{3}\, \left(\frac{1}{2} \hat{\phi}^{\prime\, 2} + \tilde{V}(\hat{\phi})\right), 
\eea
where ${^{\prime}H}=H \, a$, ${^{\prime}\phi}=\dot{\phi}\, a$ , $\tilde{G}=2\, f\, G/(3\, \lambda -1)$, the background scalar field $\hat{\phi}=\hat{\phi}(\eta)$ and inflation potential $\tilde{V}(\hat{\phi})=V(\hat{\phi})/f(\lambda)$
\bea
H^2 = \frac{16\, \pi\, f\, G}{3 (3 \lambda -1) }\, \left(V(\phi)+\frac{\Lambda_{\omega}}{8\, \pi\, G}\right). \label{eq:woprojct}
\eea
The right-hand side include energy densities related to the scalar field and to the cosmological constant. Thus, Eq. (\ref{eq:woprojct}) does not depend on equation of state (EoS). It is obvious that Eq. (\ref{eq:woprojct})  can be reduced to the standard Friedmann equation, i.e. GR gravity, at $f=\lambda=1$.

\section{Results}
\label{sec:Fried}

\subsection{Cosmic inflation models}
\label{sec:cim}

\begin{figure}
\centering{
\includegraphics[width=6.5cm,angle=-90]{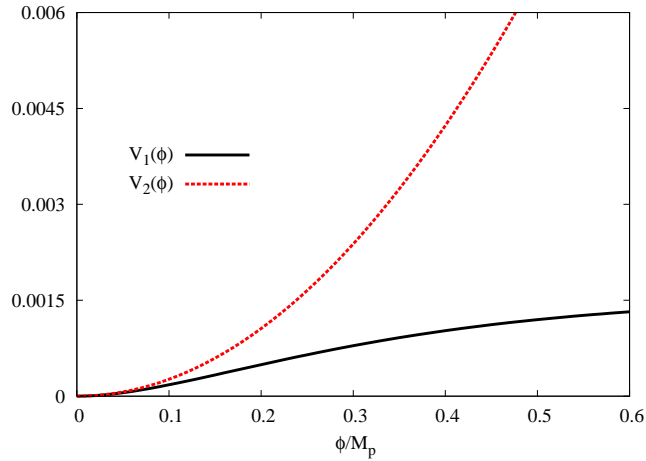}
\caption{The hybrid inflation potential $V(\phi)$ with quadratic- [Eqs. (\ref{poweri}) solid curve] and higher-term [(\ref{eq:mssm}) dashed curve] is given in dependence on varying  inflation field $\phi/M_{pl}$. 
\label{pot}
}}
\end{figure}

We propose to implement two models for the inflation potential:
\begin{enumerate}
\item  
for a power-law inflation with the free parameter $d$, an inflation potential is given as \cite{Starobinsky,Liddle:1993}
\bea
V_1 (\phi)=\frac{3 M_{pl}^2 d^2}{32 \pi} \left[1-\exp \left(-\frac{16\pi}{3 M_{pl}^2 }^{1/2} \phi \right) \right]^2, \label{poweri}
\eea

\item
and another inflation potential is based on certain minimal supersymmetric extensions of the standard model for elementary particles \cite{allahverdi-2006}
\bea
V_2(\phi) = \left(\frac{m^2}{2}\right)\,\phi^2   - \left(\frac{\sqrt{2\,\lambda\,(n-1)}\,m}{n}\right)\, \phi^n  + \left(\frac{\lambda}{4}\right)\,\phi^{2(n-1)}.
\eea
The effects of this scalar potential have been studied, recently~\cite{allahverdi-2006,sanchez-2007}. There are two free parameters, $m$ and $\lambda$, while $n>2$ is an integer. At $n=3$, 
 \bea
V_2(\phi) = \left(\frac{m^2}{2}\right)\,\phi^2   - \left(\frac{2\sqrt{\lambda}\,m}{3}\right)\, \phi^3  + \left(\frac{\lambda}{4}\right)\,\phi^{4}, \label{eq:mssm}
\eea
This approximation shall be utilized in the present calculations.
\end{enumerate}

The dependence of the hybrid inflation with quadratic- and higher-term potential ($V(\phi)$) on varying inflation field ($\phi/M_{pl}$), Eqs.  (\ref{poweri}) and (\ref{eq:mssm}), respectively, is illustrated in Fig. \ref{pot}. We find that $V_1(\phi)$ slowly increases with increasing $\phi/M_{pl}$ (solid curve).  It can approach a saturation at large $\phi/M_{pl}$, while $V_2(\phi)$ shows a rapid and apparently steady increase (dashed curve). Bearing in mind these differences, it intends to compare between the impacts of these scalar fields on the inflationary regime.

\subsection{Fluctuations and Slow-Roll Parameters} 
\label{FluctuationHLG}

\begin{figure}
\centering{
\includegraphics[width=5.5cm,angle=-90]{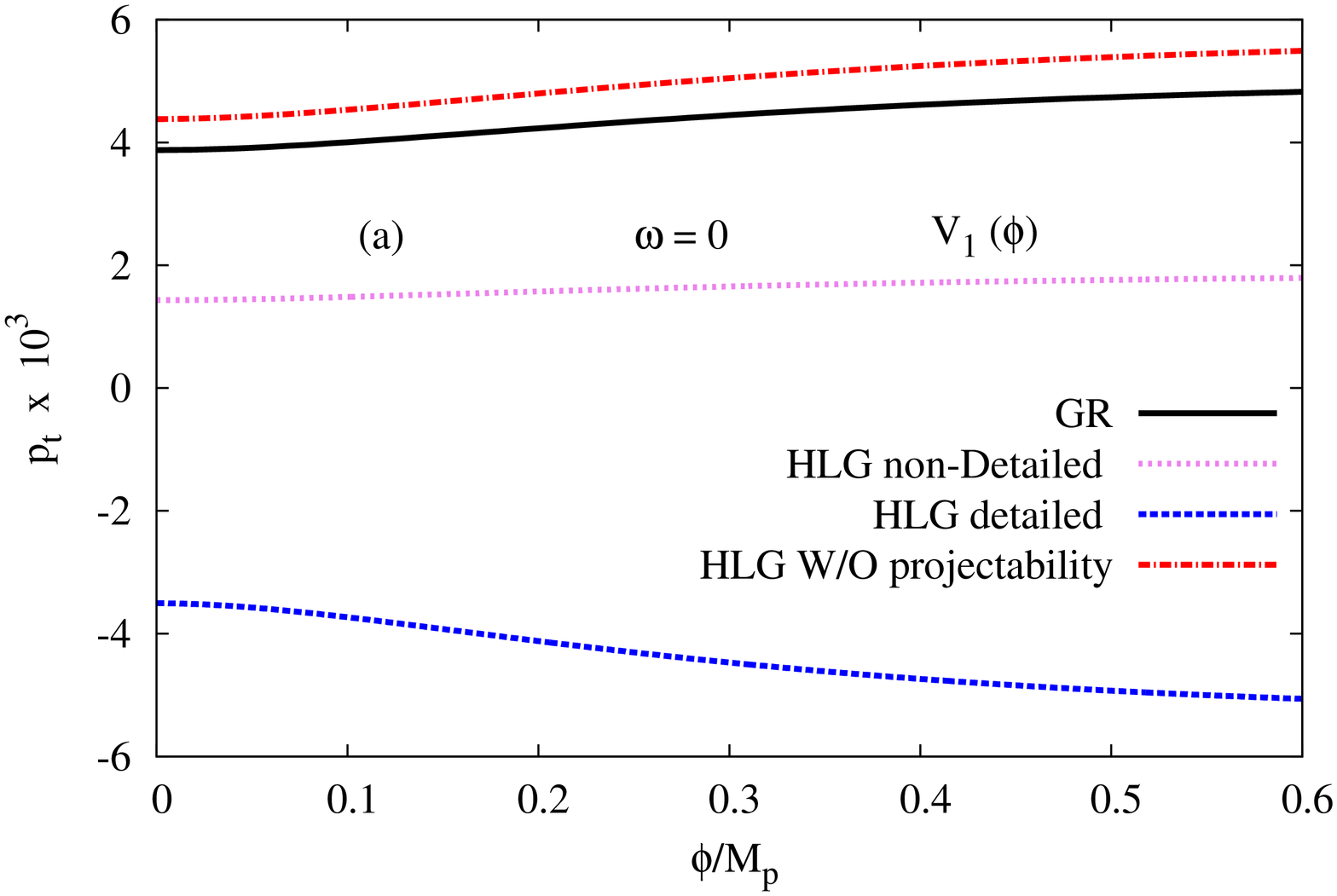}
\includegraphics[width=5.5cm,angle=-90]{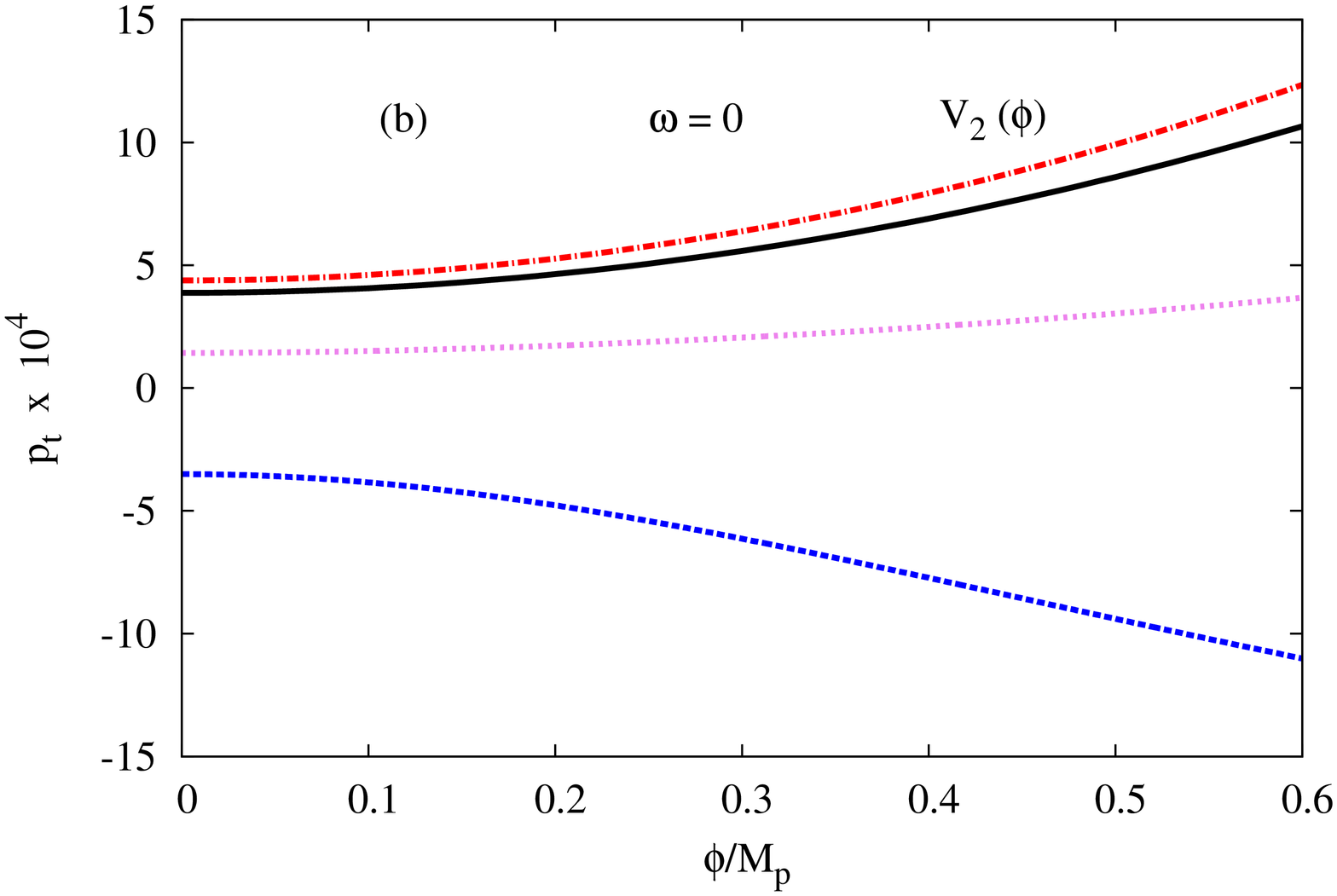}\\
\includegraphics[width=5.5cm,angle=-90]{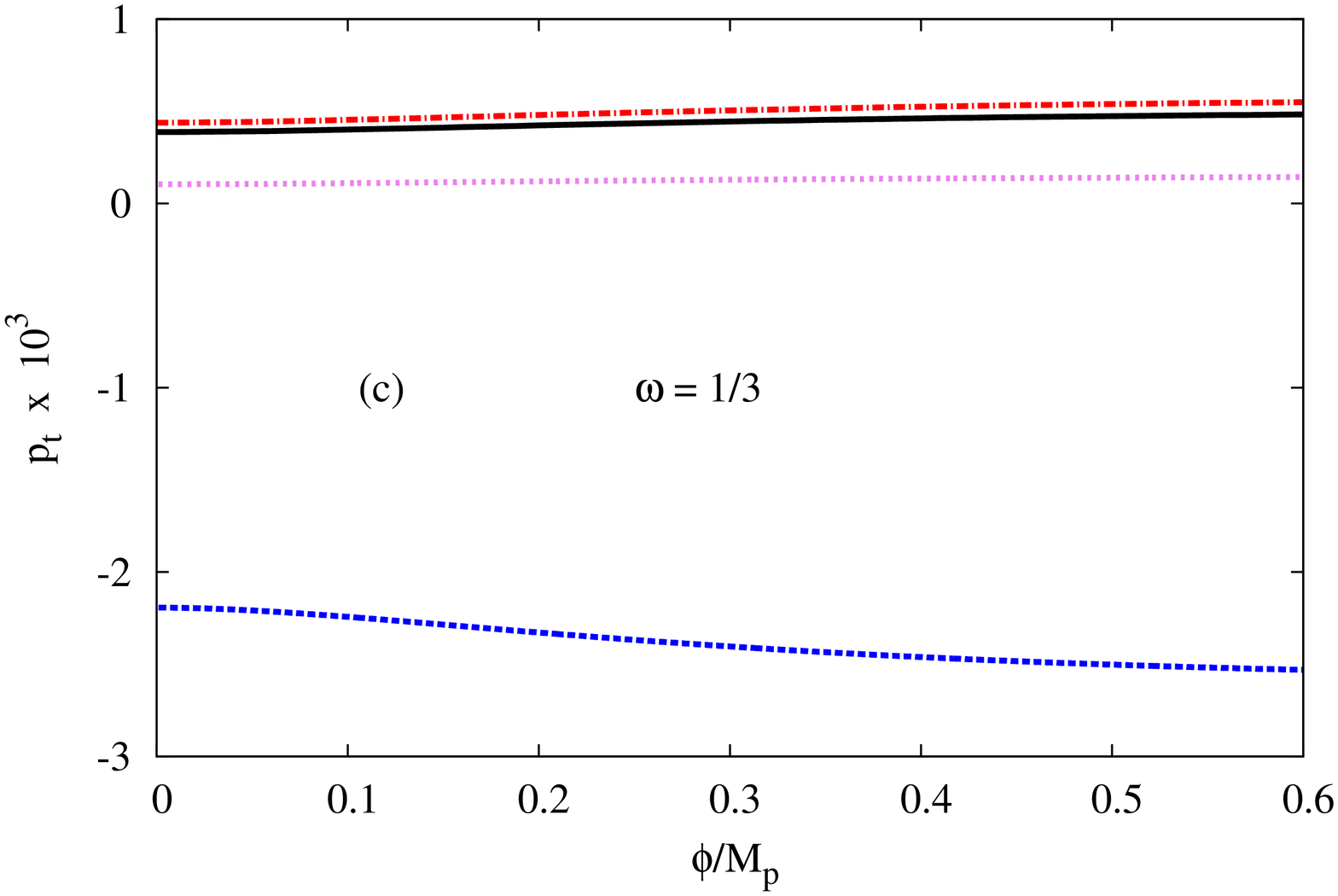}
\includegraphics[width=5.5cm,angle=-90]{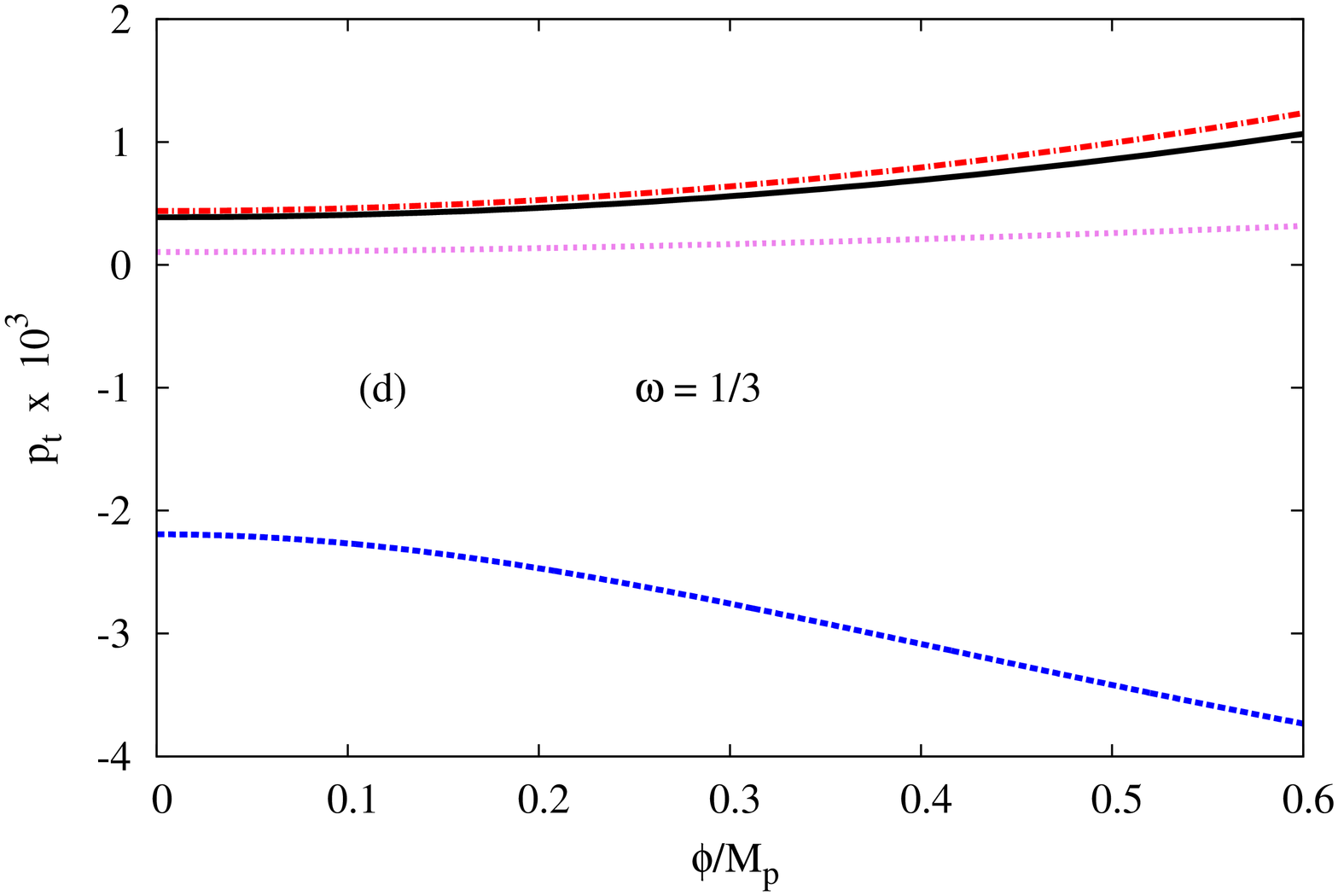}\\
\includegraphics[width=5.5cm,angle=-90]{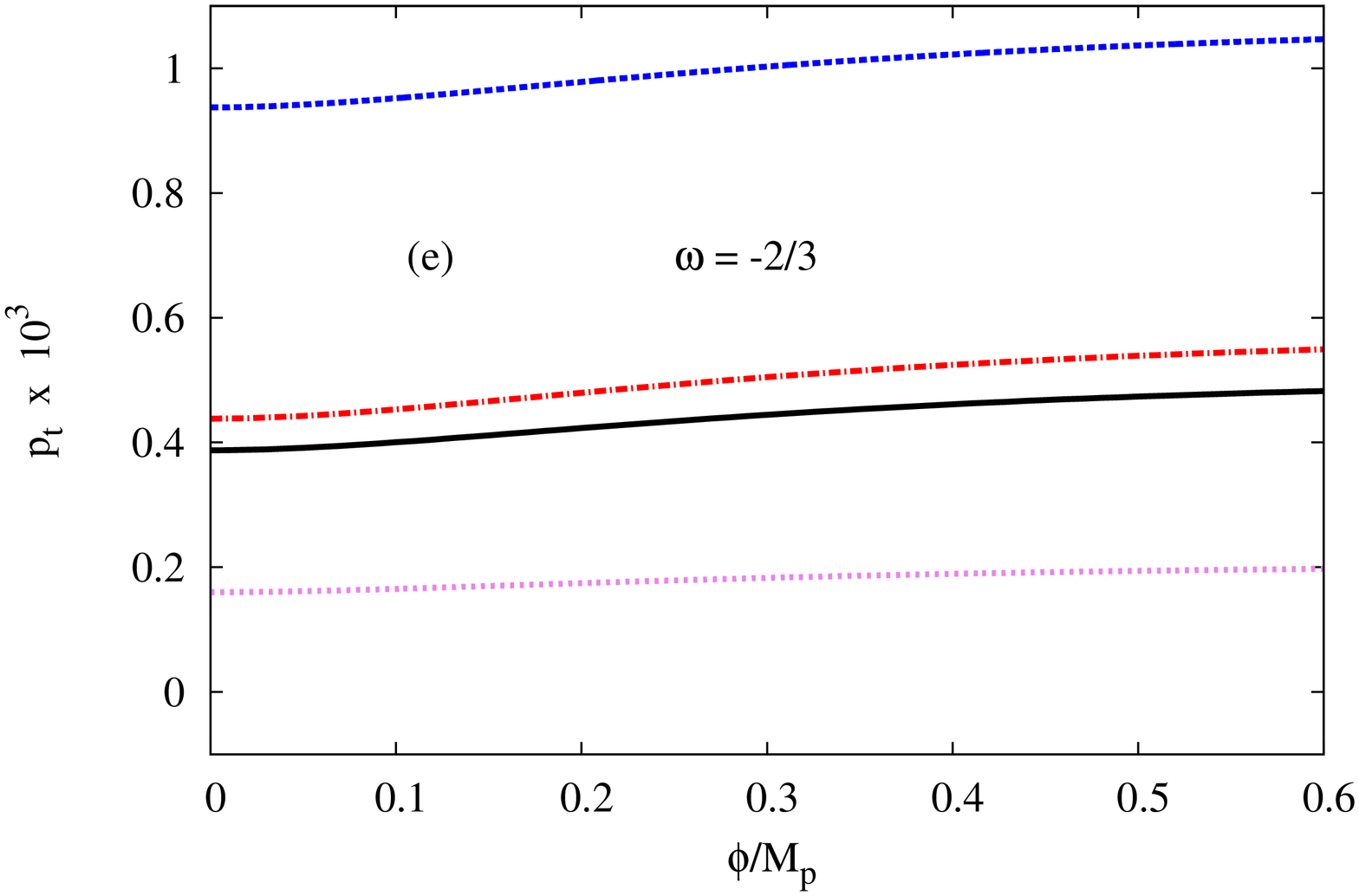}
\includegraphics[width=5.5cm,angle=-90]{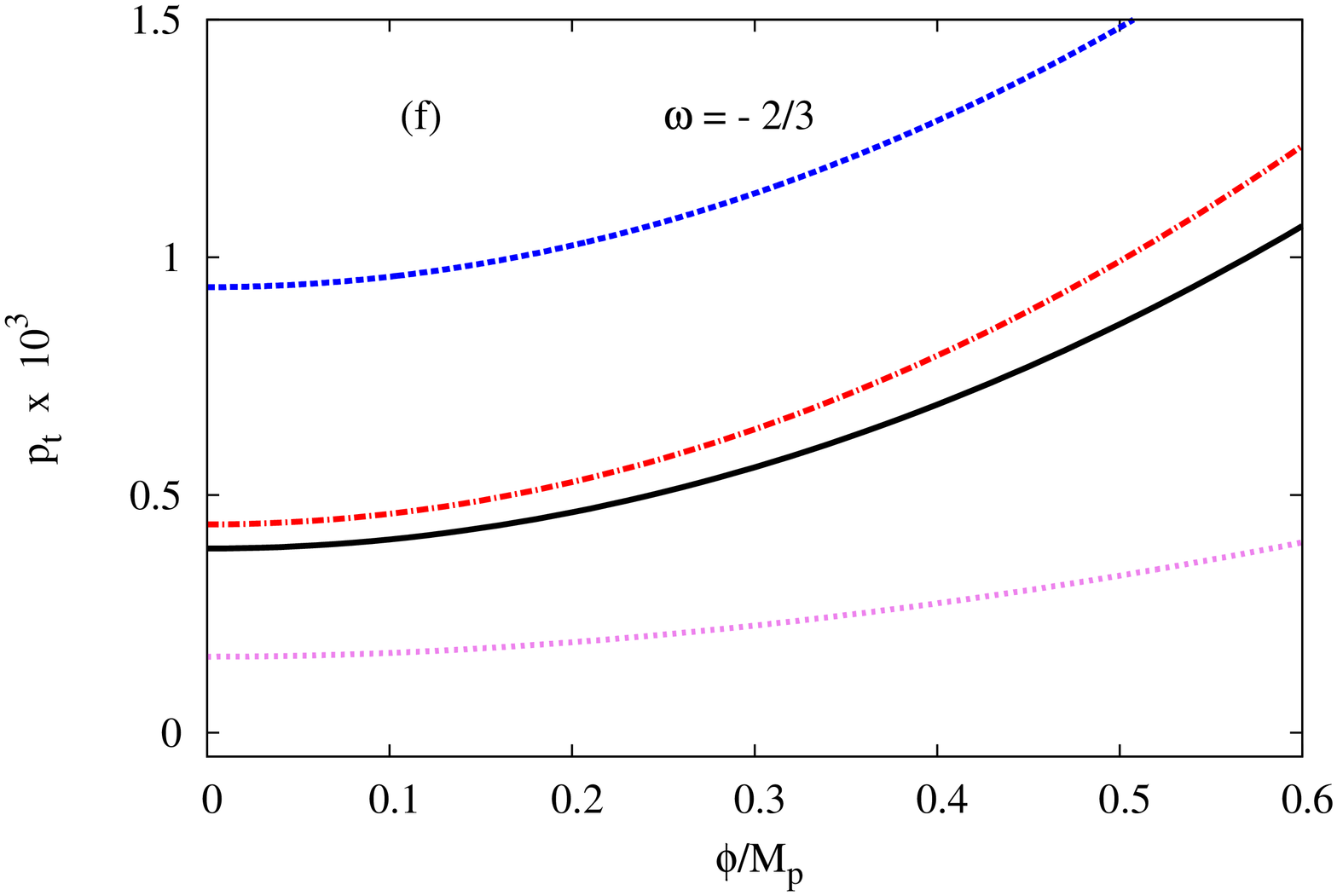}
\caption{The tonsorial density fluctuations ($P_t$) from GR (solid curve), HL gravity detailed (dashed curve), HL gravity non-detailed (dotted curve) and HL gravity without the projectability (dash-dotted curve) are given as functions of the inflation field ($\phi$) in units of Planck mass $M_{p}$ for two inflation potentials and various EoS; $\omega =0$ (top panel), $1/3$ (middle panel), and $-2/3$ (bottom panel).
\label{tensorialfluc}
}}
\end{figure}

\begin{figure}
\centering{
\includegraphics[width=5.5cm,angle=-90]{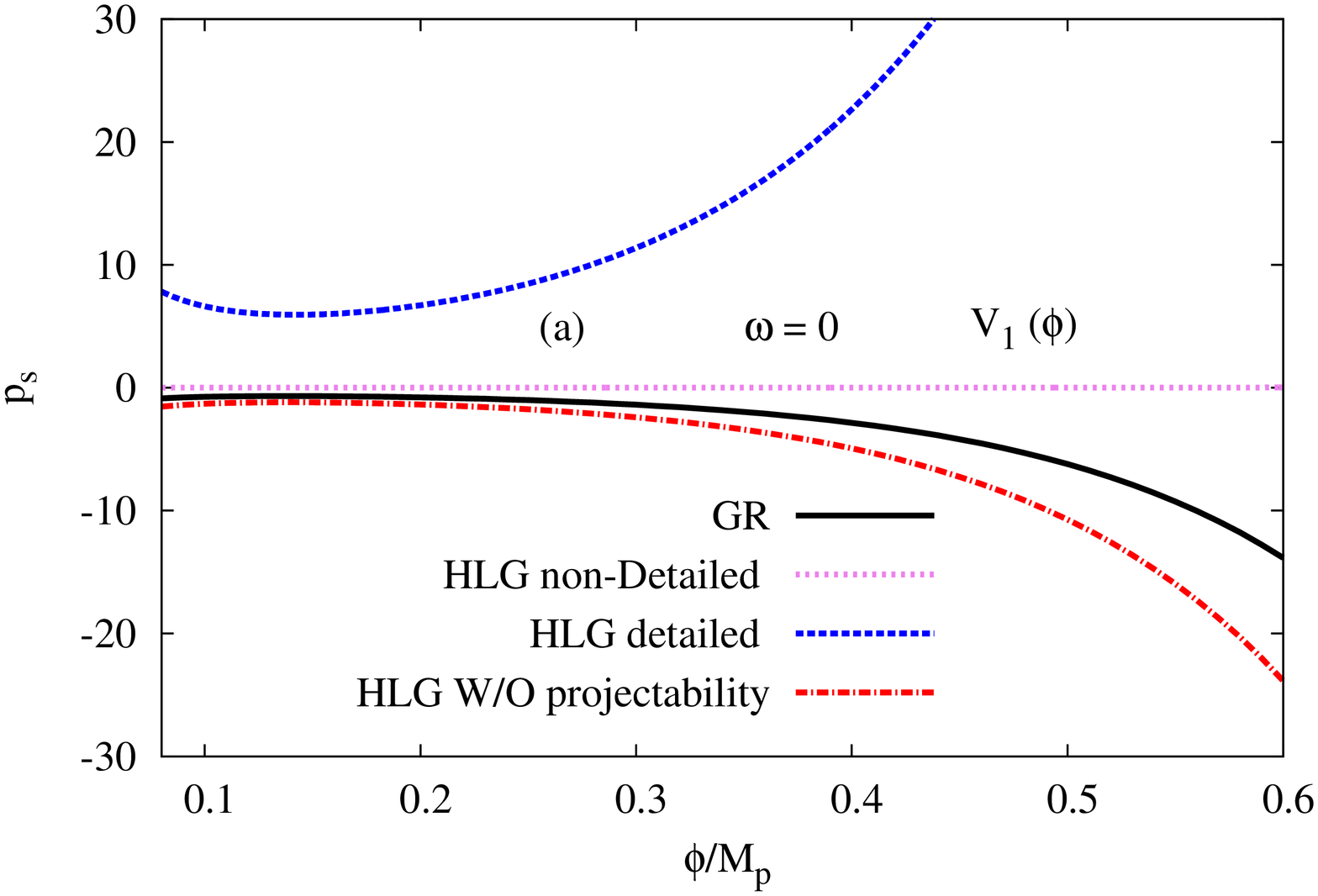}
\includegraphics[width=5.5cm,angle=-90]{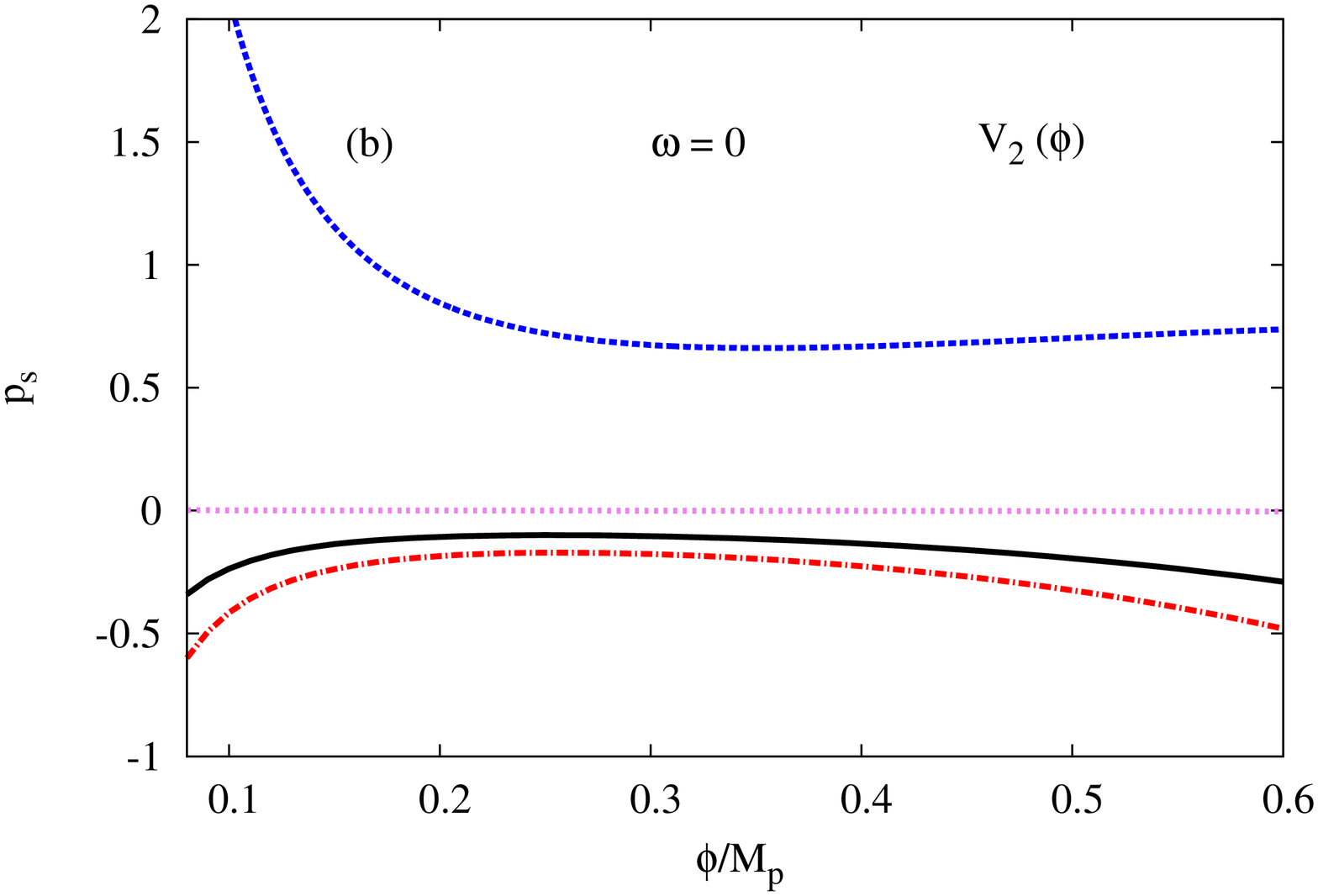}\\
\includegraphics[width=5.5cm,angle=-90]{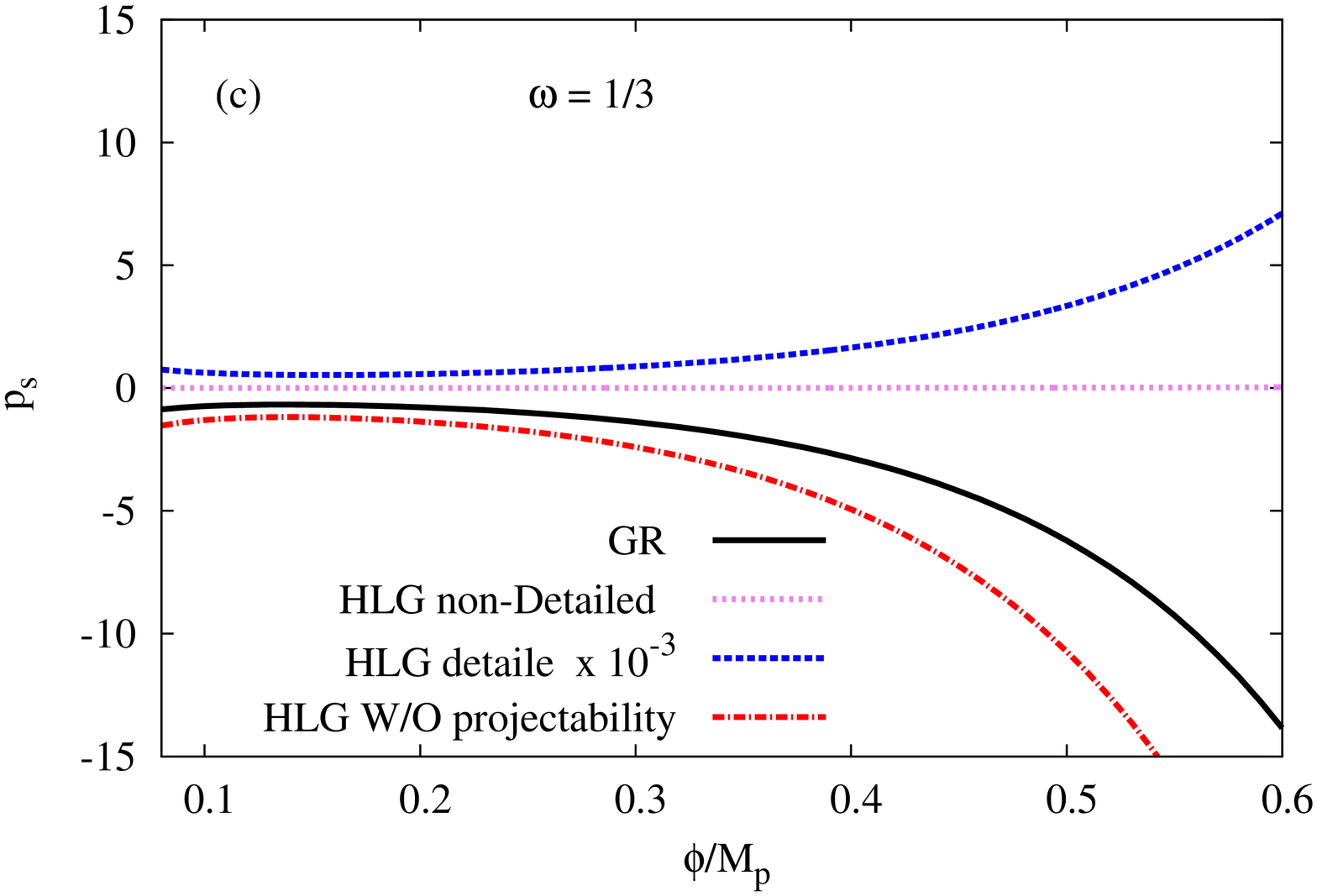}
\includegraphics[width=5.5cm,angle=-90]{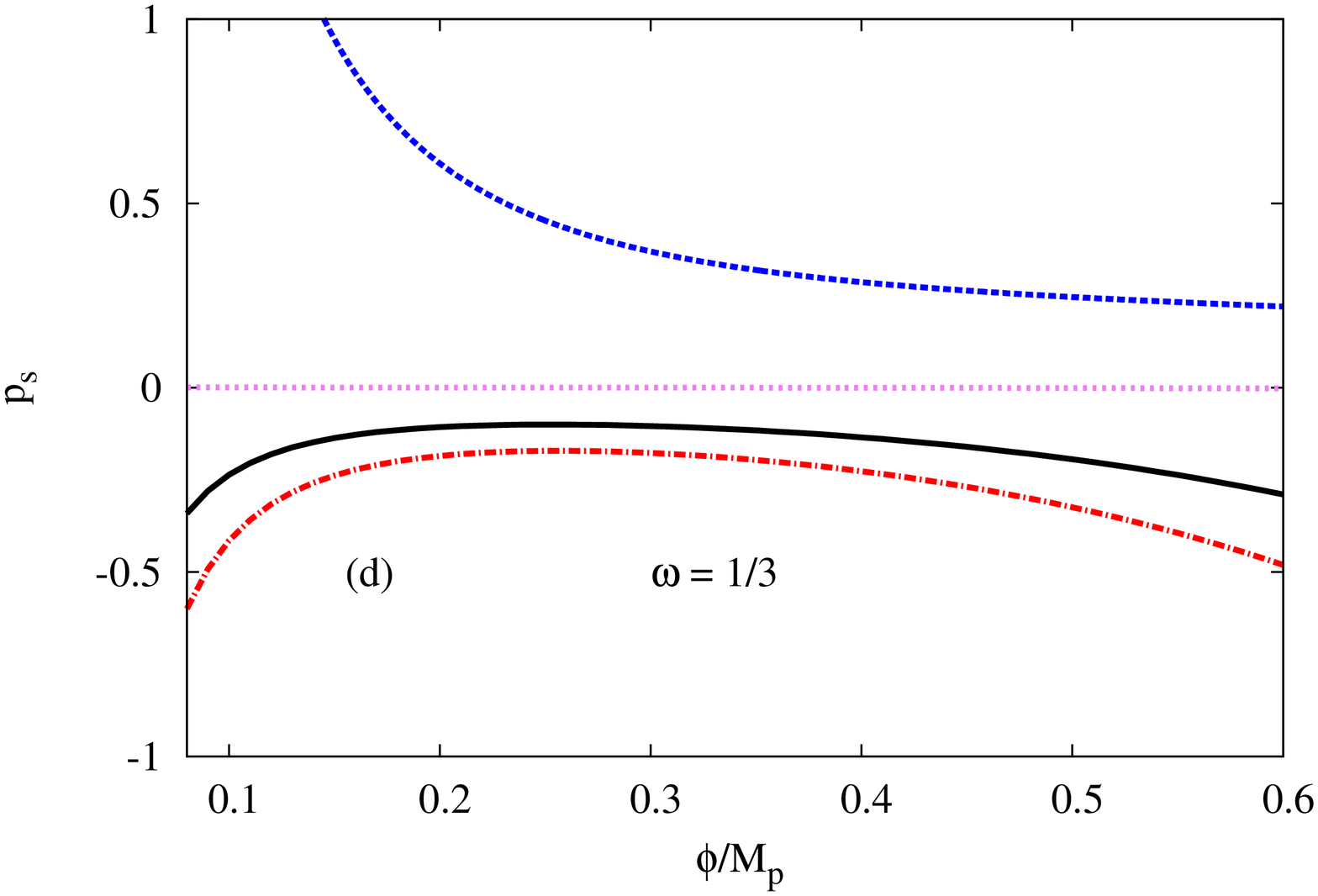}\\
\includegraphics[width=5.5cm,angle=-90]{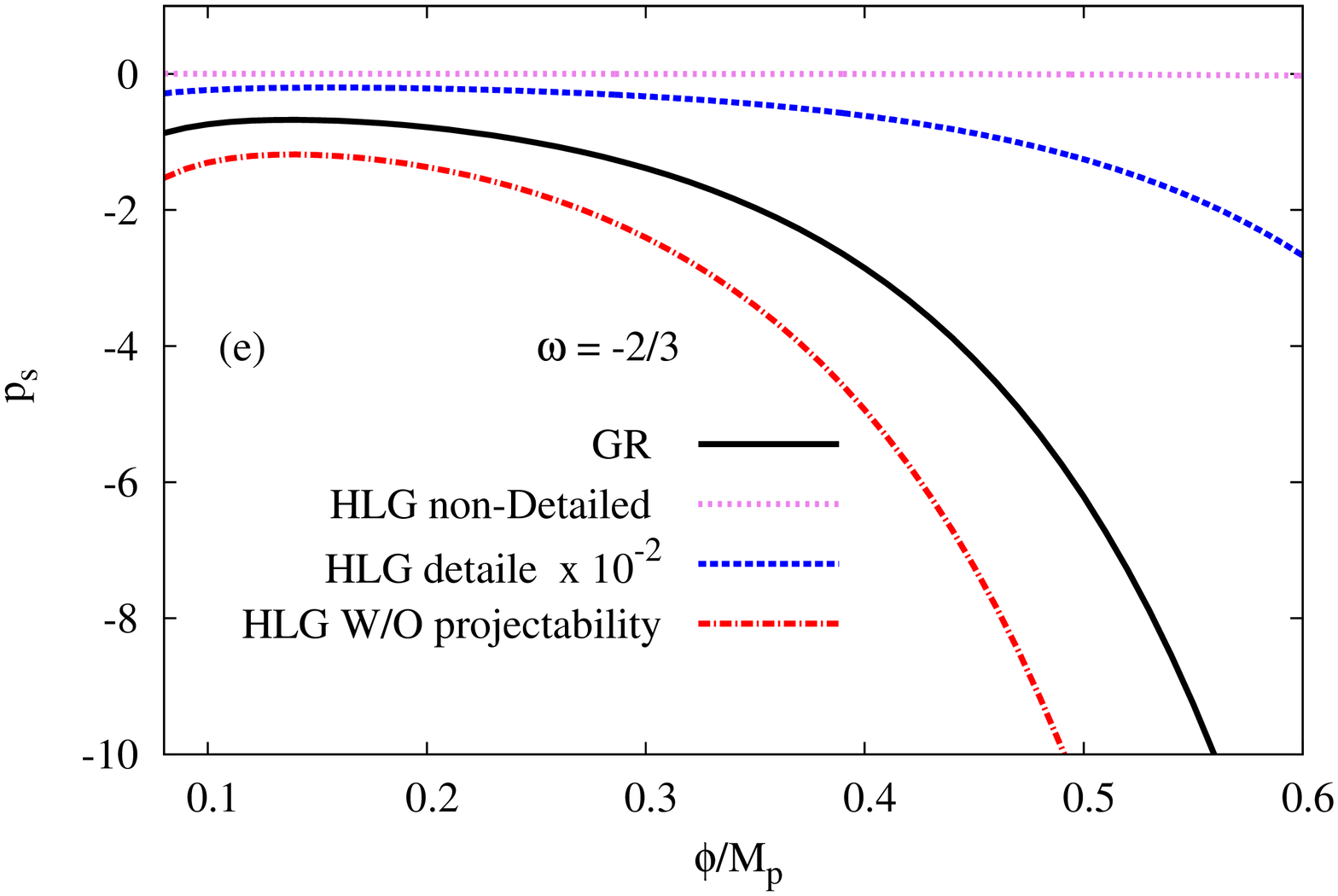}
\includegraphics[width=5.5cm,angle=-90]{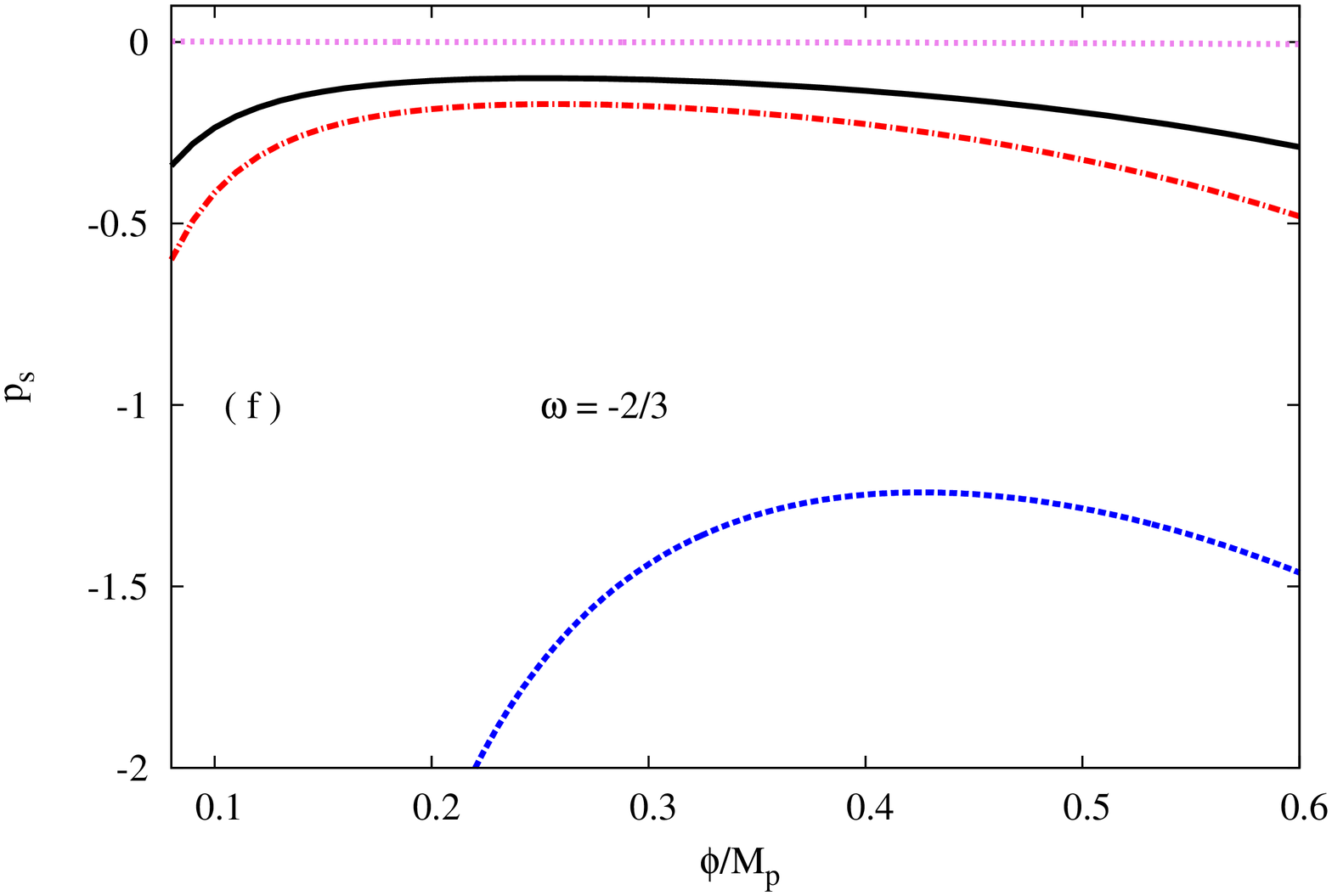}\\
\caption{The same as in Fig. \ref{tensorialfluc} but for spectral density fluctuations ($P_s$).
\label{spectralfluc}
}}
\end{figure}

In very early Universe, the scaler field ($\phi$) is assumed to derive the inflation  \cite{Linde:1982,Liddle:1993,Liddle:1995}.  As introduced in Ref. \cite{0904.2190}, it was believed that the detailed balance conditions for HL gravity were not necessary for the inflation. But, serious horizon problems, such as monopole and domain walls, remain unsolved. The inclusion of scalar fields is essential to generate inflation even in HL gravity. The main  slow-roll parameters are given as
\bea
\epsilon _H  &\equiv & \frac{M_{pl}^2}{16 \, \pi} \left(\frac{\partial _{\phi} V(\phi)}{ V(\phi)}\right)^2,
\label{paramters1} \\
\eta _H & \equiv & \frac{M_{pl}^2}{8 \pi} \left(\frac{\partial_{\phi}^2 V(\phi)}{V(\phi)}\right),
\label{paramters2}
\eea
where $V(\phi)$ can be any inflation potential such as either Eq. (\ref{poweri}) or (\ref{eq:mssm}). The main reason why this scalar field or another one is chosen is obviously merely motivated by its ability to cope with trusted observations. We shall judge about the different fields due their abilities to reproduce the recent PLANCK observations. We utilize different equations of state characterizing the cosmic background geometry.

\begin{figure}
\centering{
\includegraphics[width=5.5cm,angle=-90]{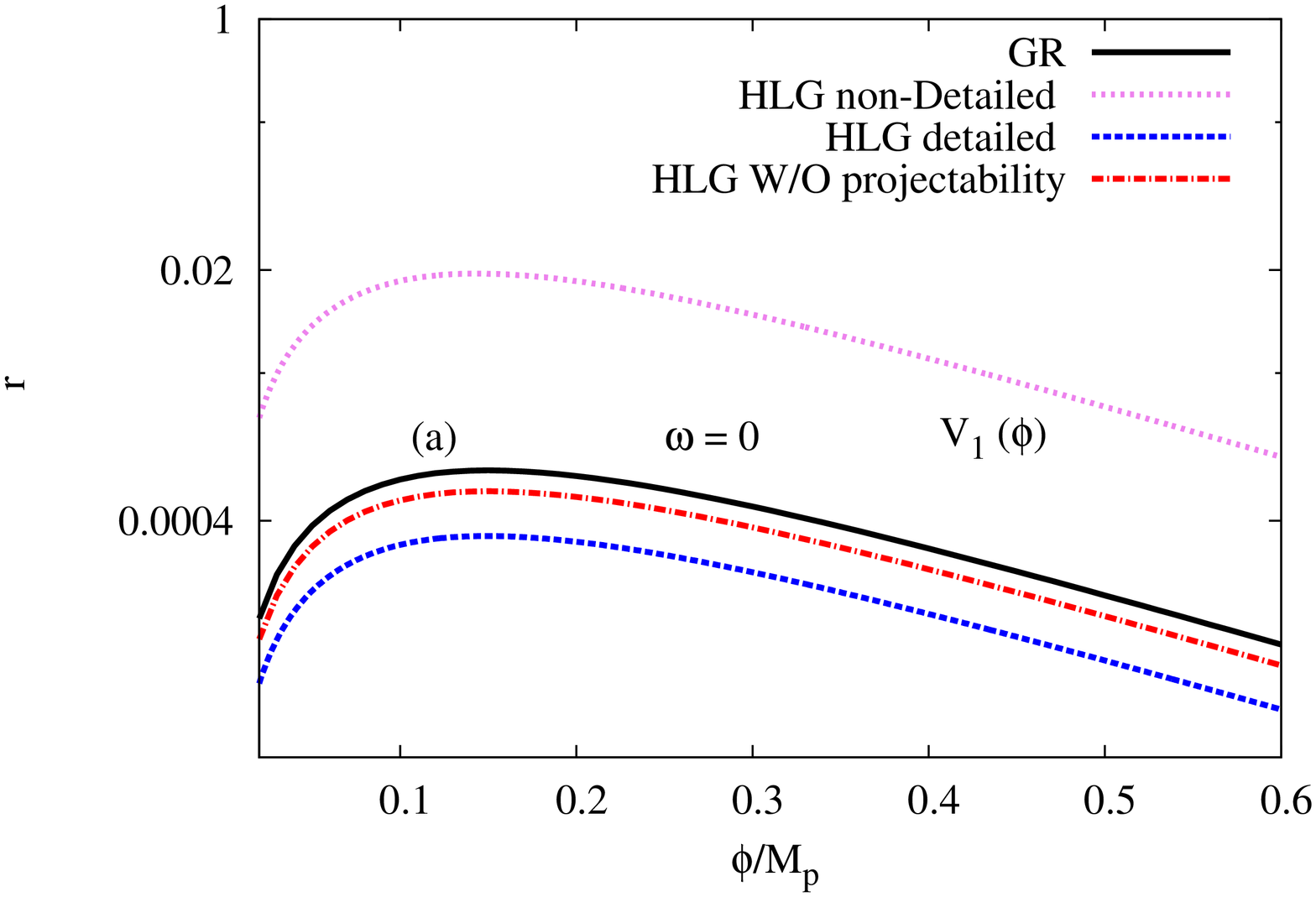}
\includegraphics[width=5.5cm,angle=-90]{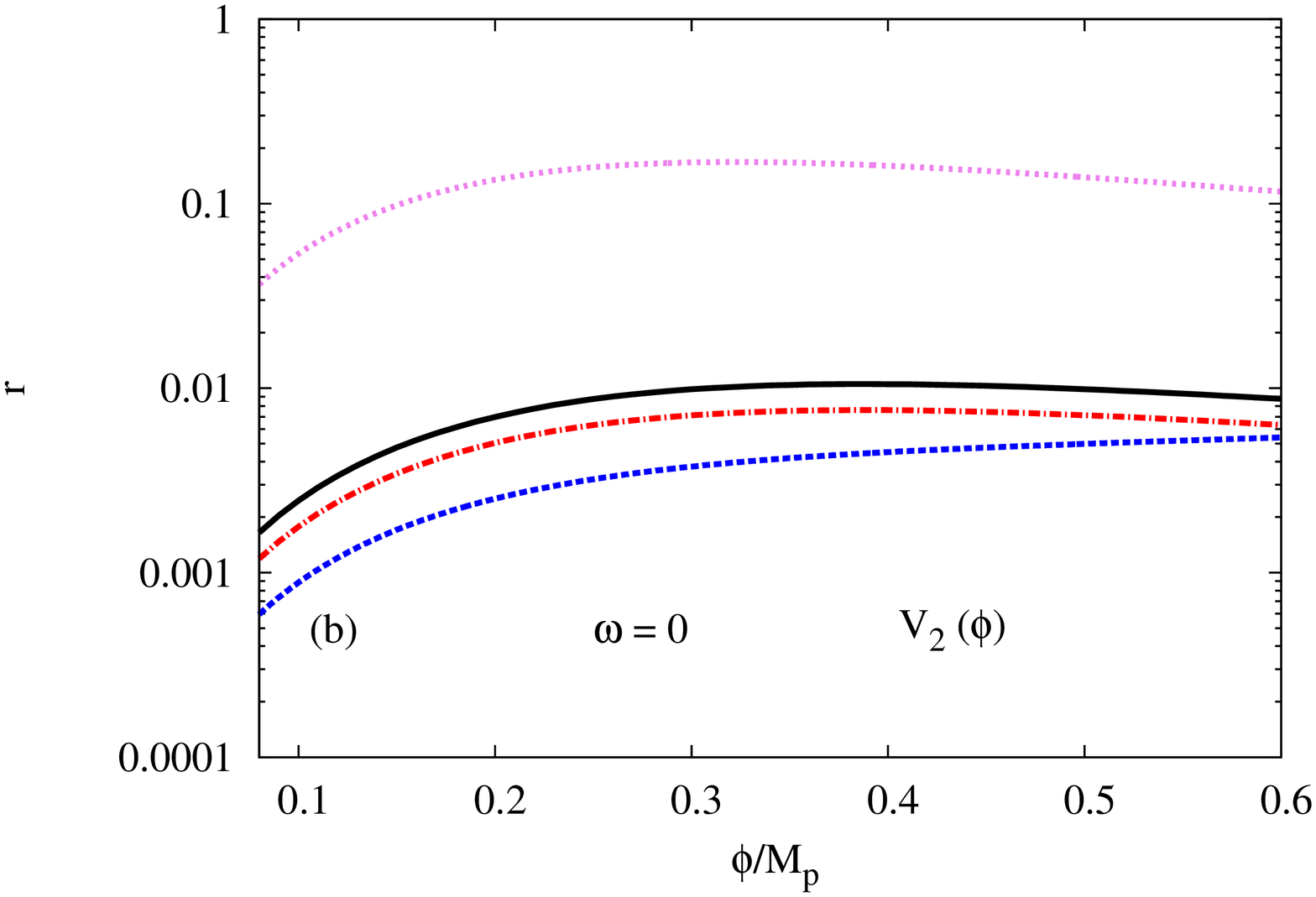}\\
\includegraphics[width=5.5cm,angle=-90]{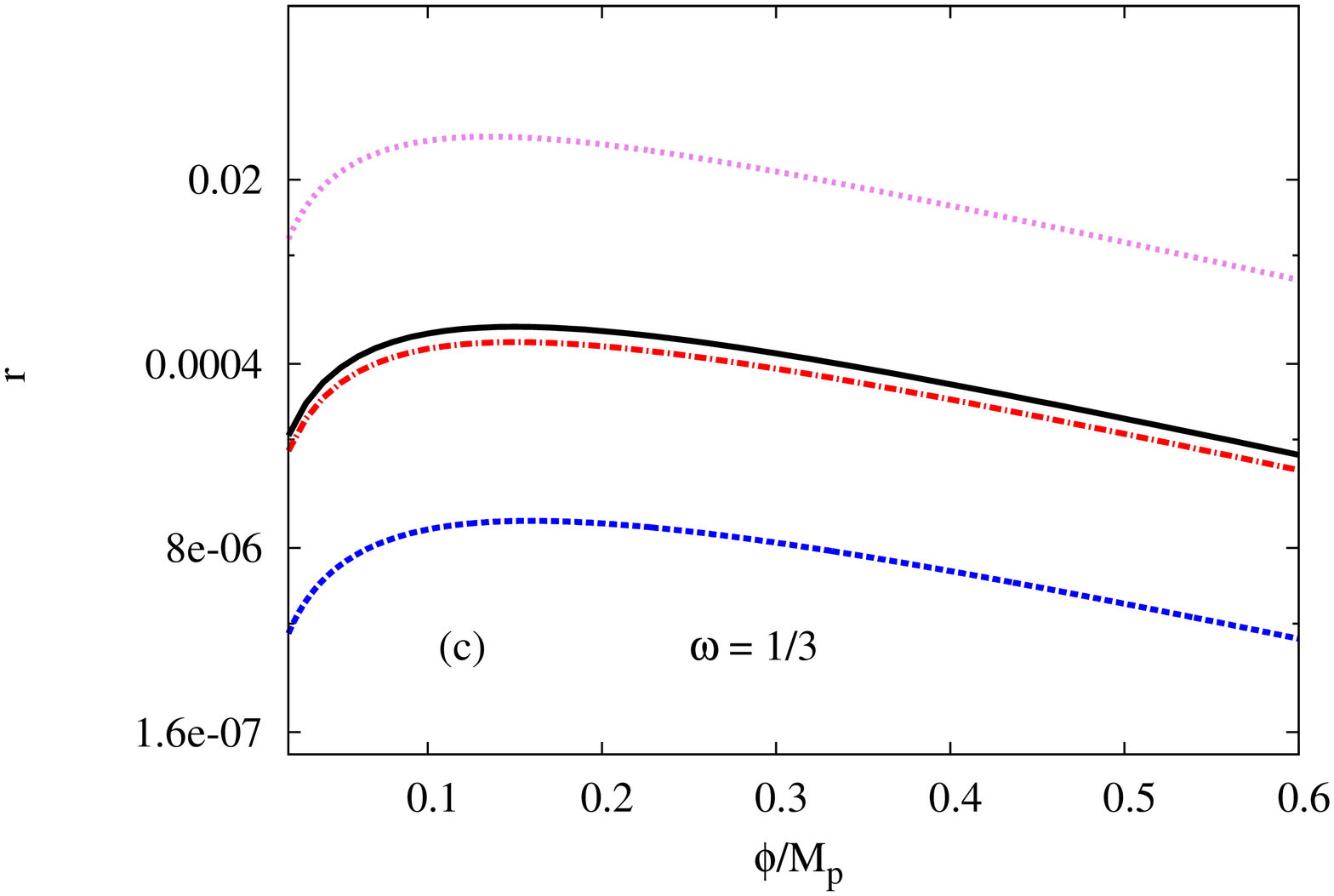}
\includegraphics[width=5.5cm,angle=-90]{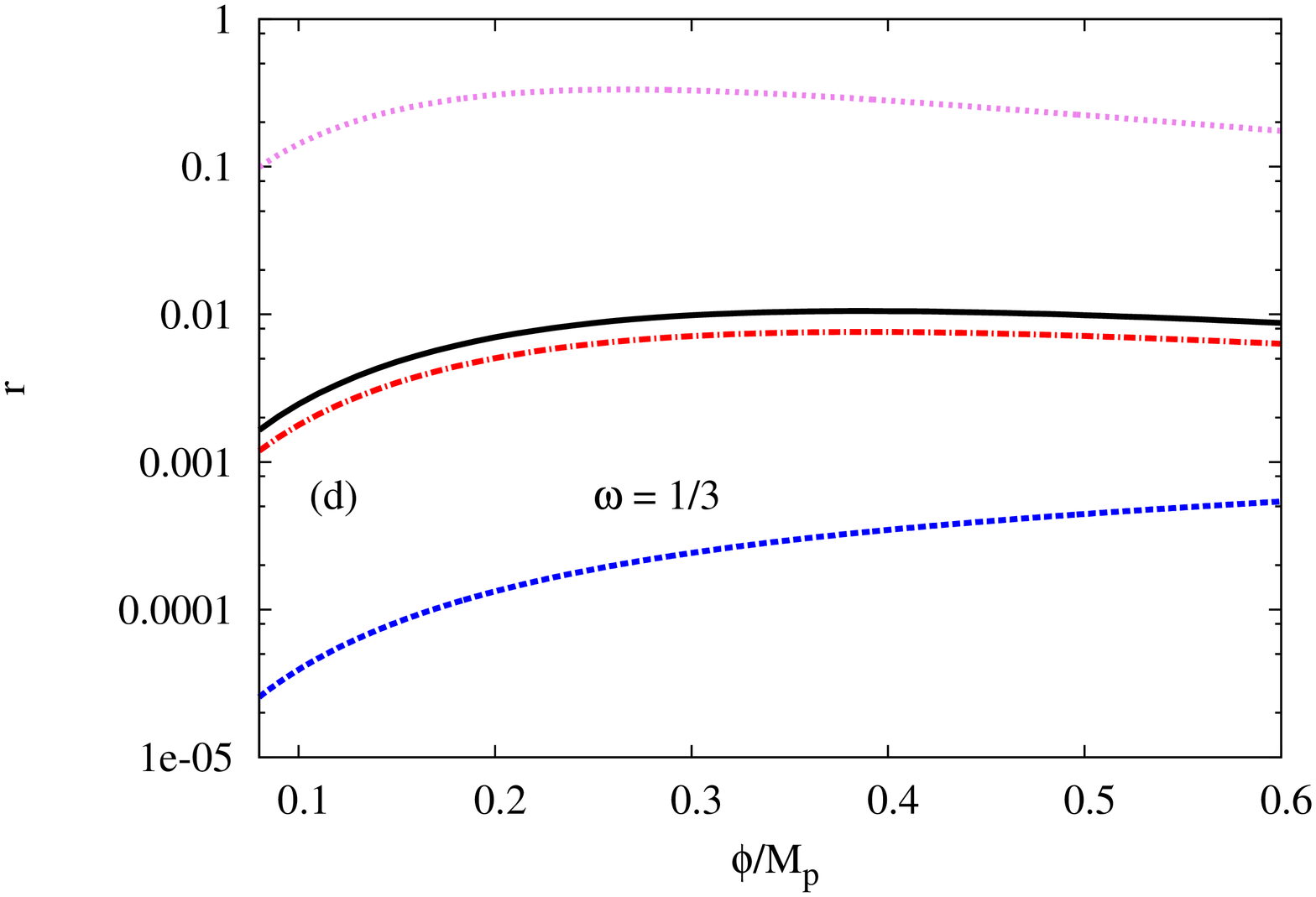}\\
\includegraphics[width=5.5cm,angle=-90]{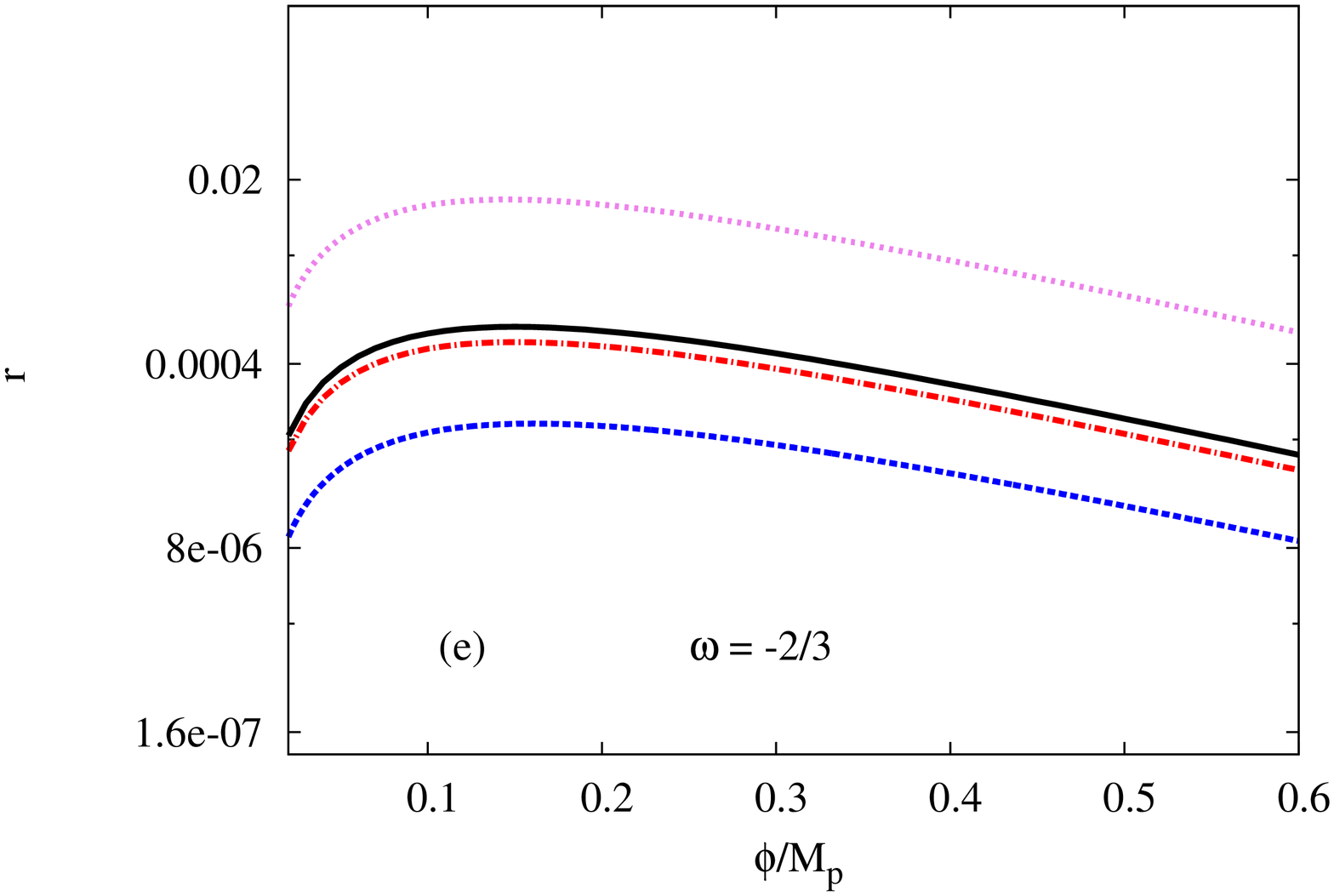}
\includegraphics[width=5.5cm,angle=-90]{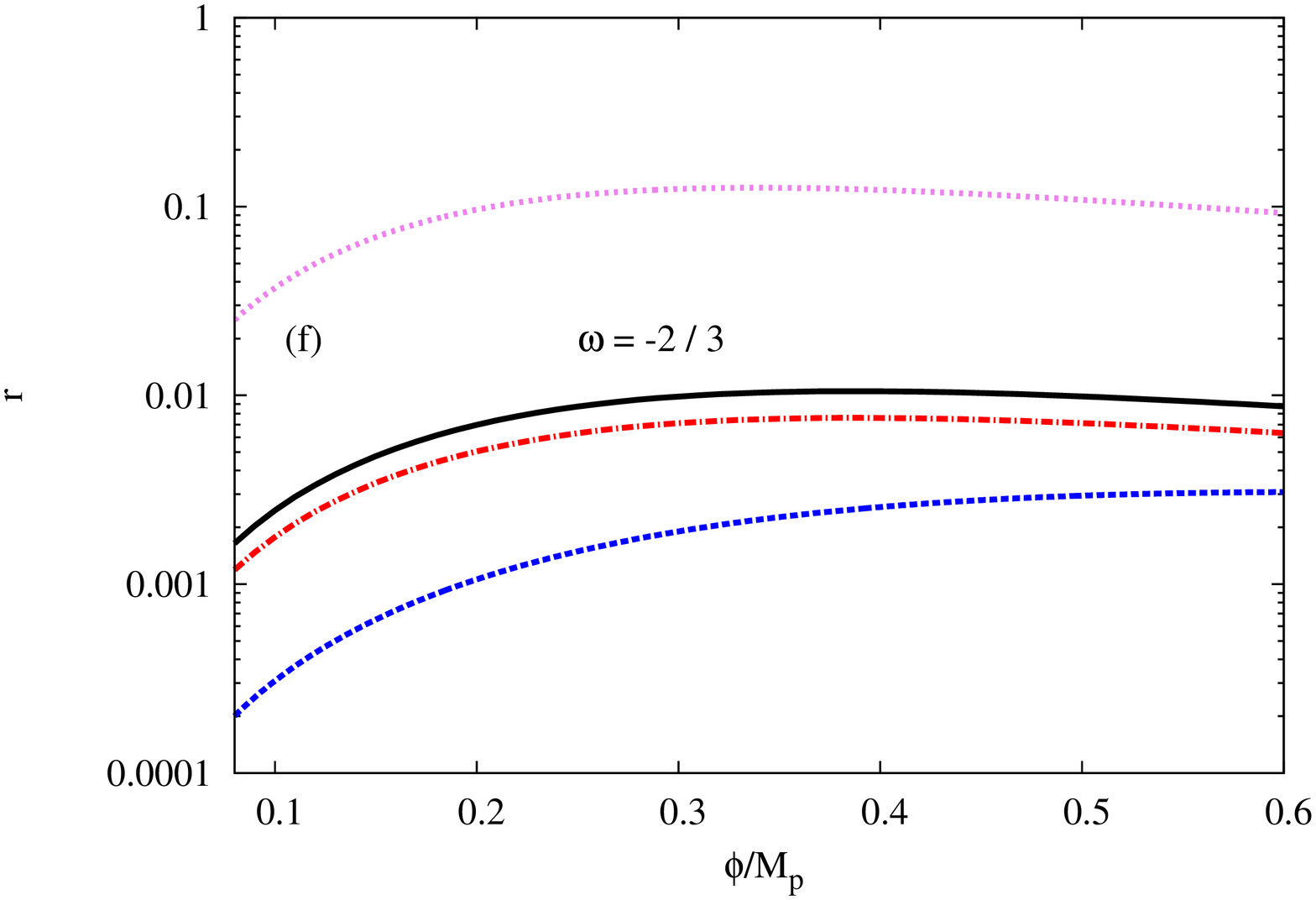}\\
\caption{The same as Fig. \ref{tensorialfluc} but for tonsorial-to-spectral density fluctuations ($r=P_t/P_s$).
\label{tensorialspectralfluc}
}}
\end{figure}

\begin{figure}
\centering{
\includegraphics[width=5.5cm,angle=-90]{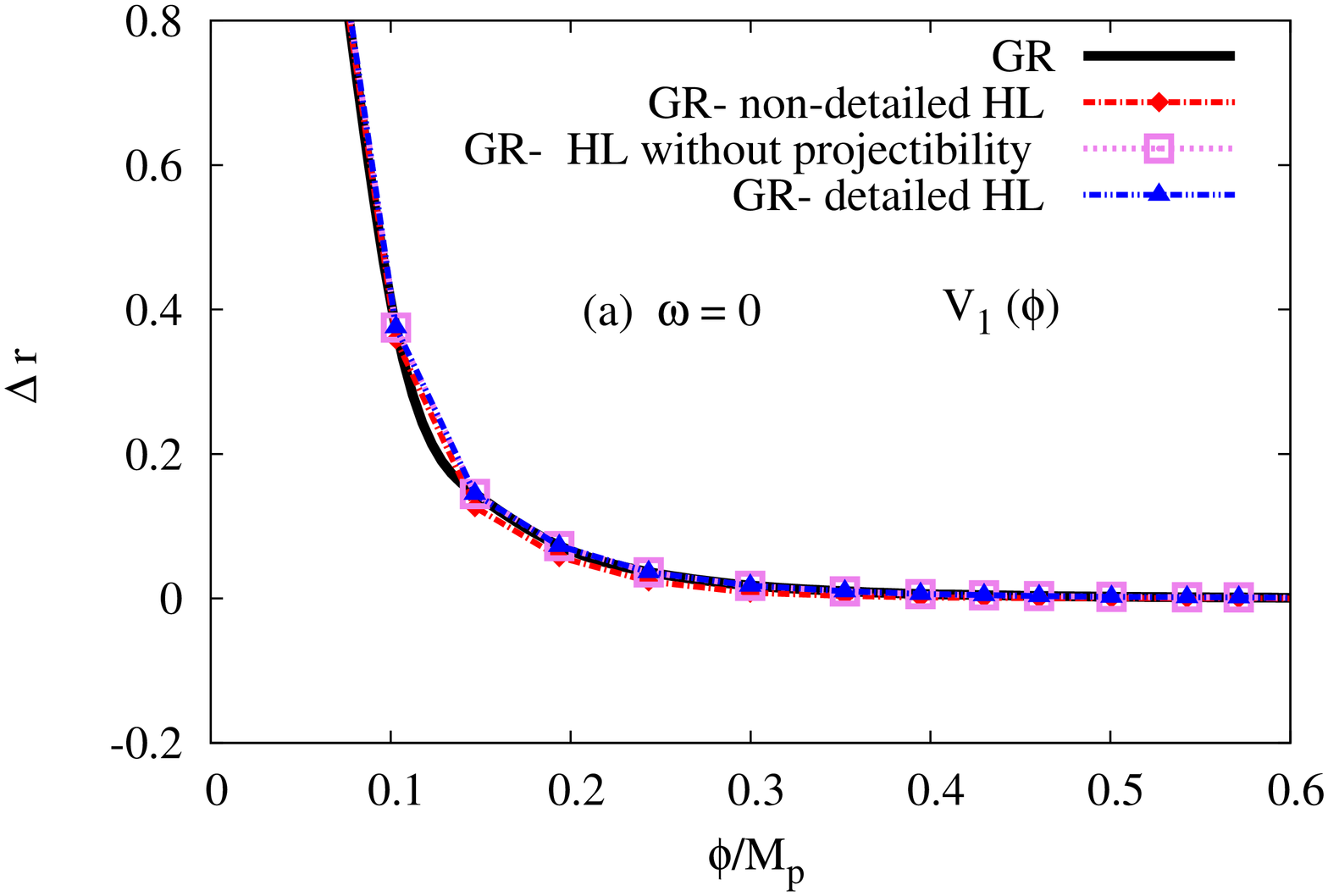}
\includegraphics[width=5.5cm,angle=-90]{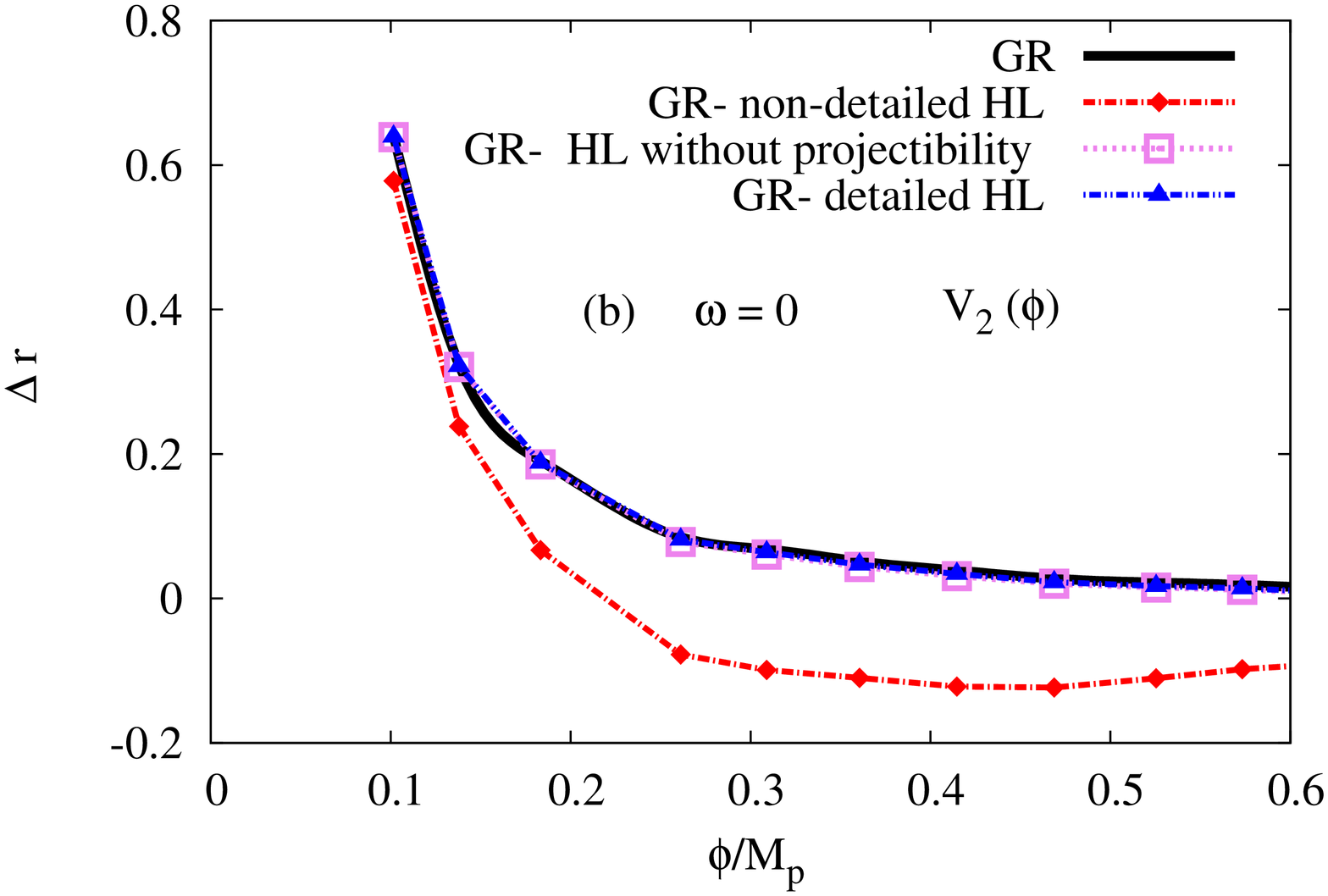}\\
\includegraphics[width=5.5cm,angle=-90]{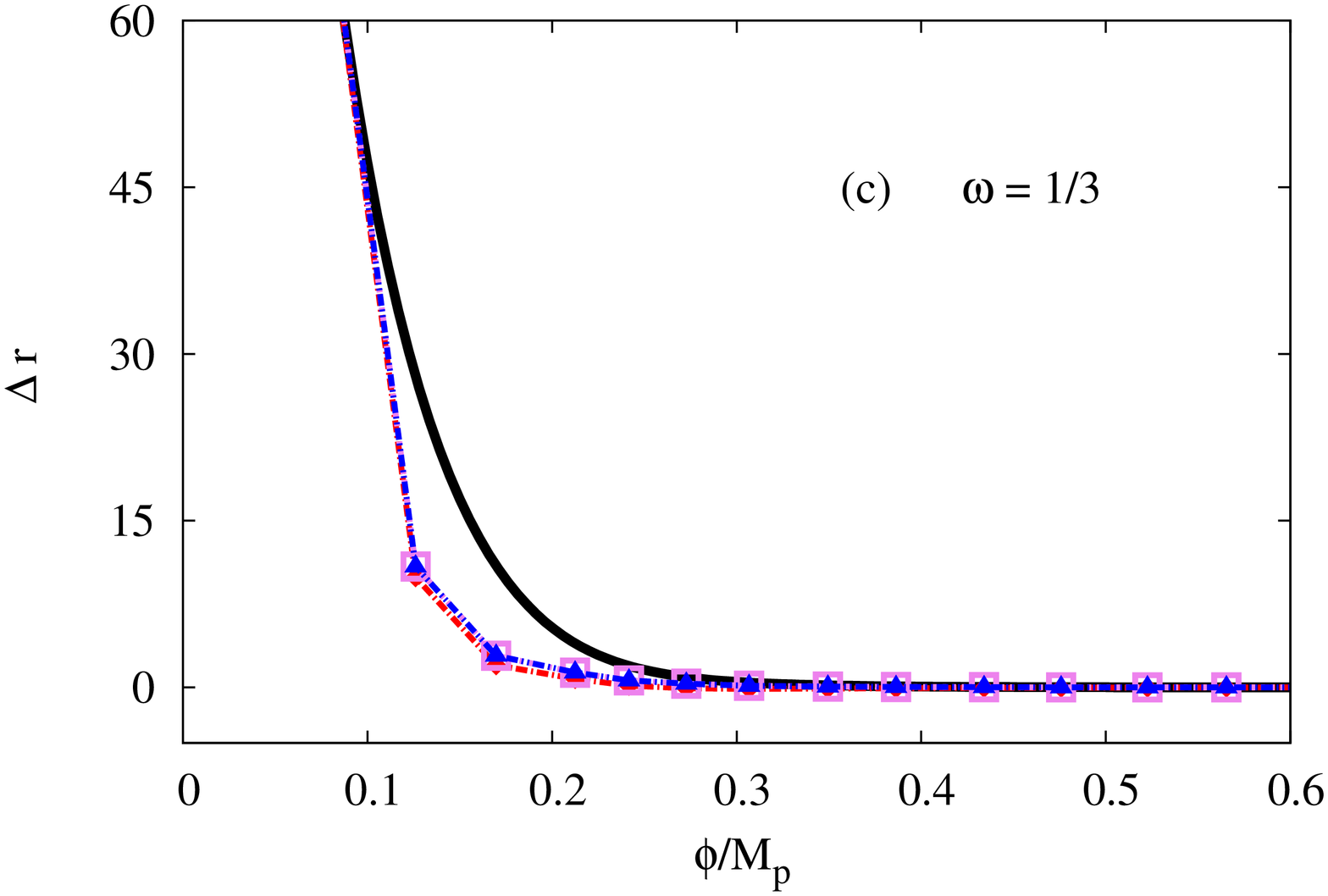}
\includegraphics[width=5.5cm,angle=-90]{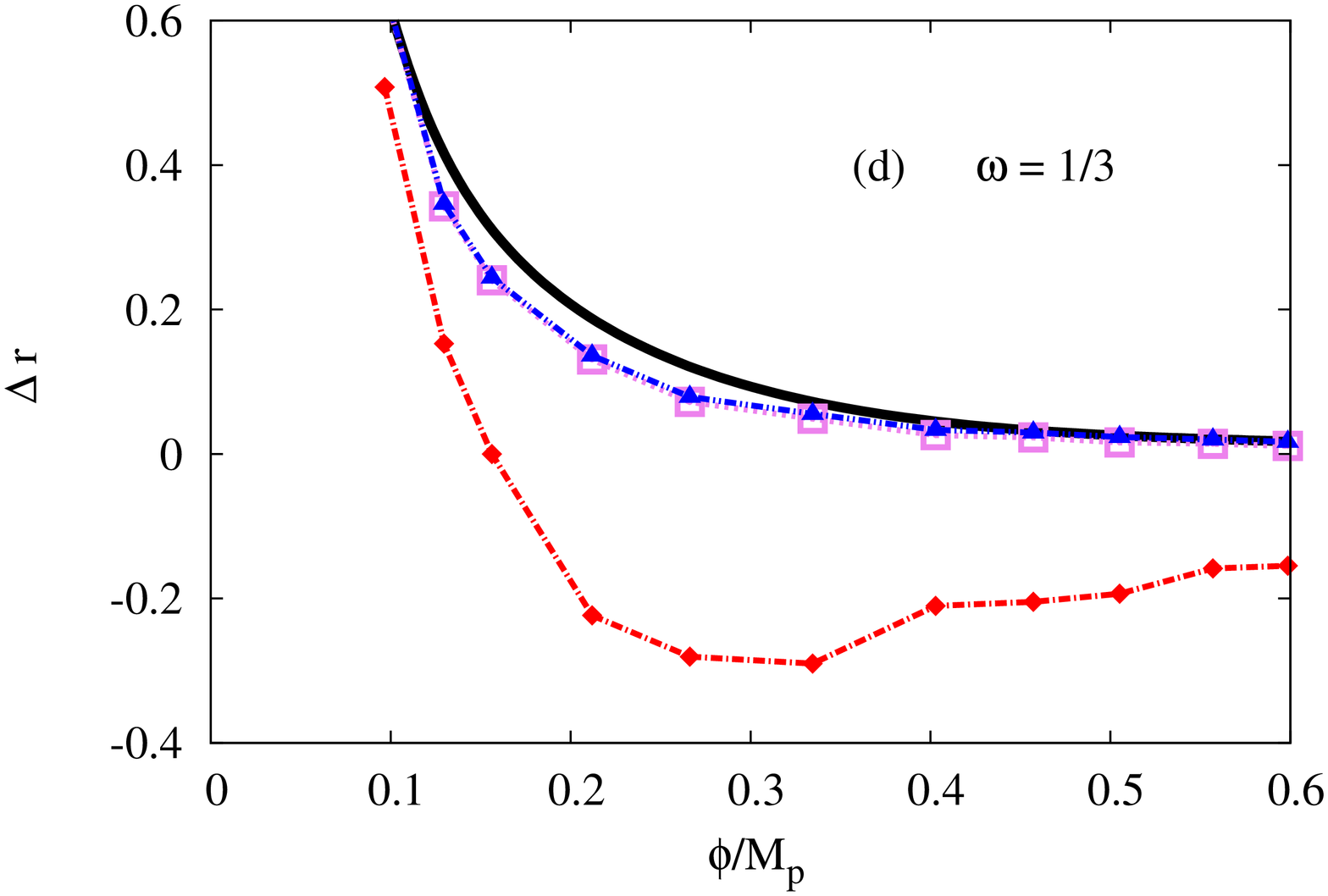}\\
\includegraphics[width=5.8cm,angle=-90]{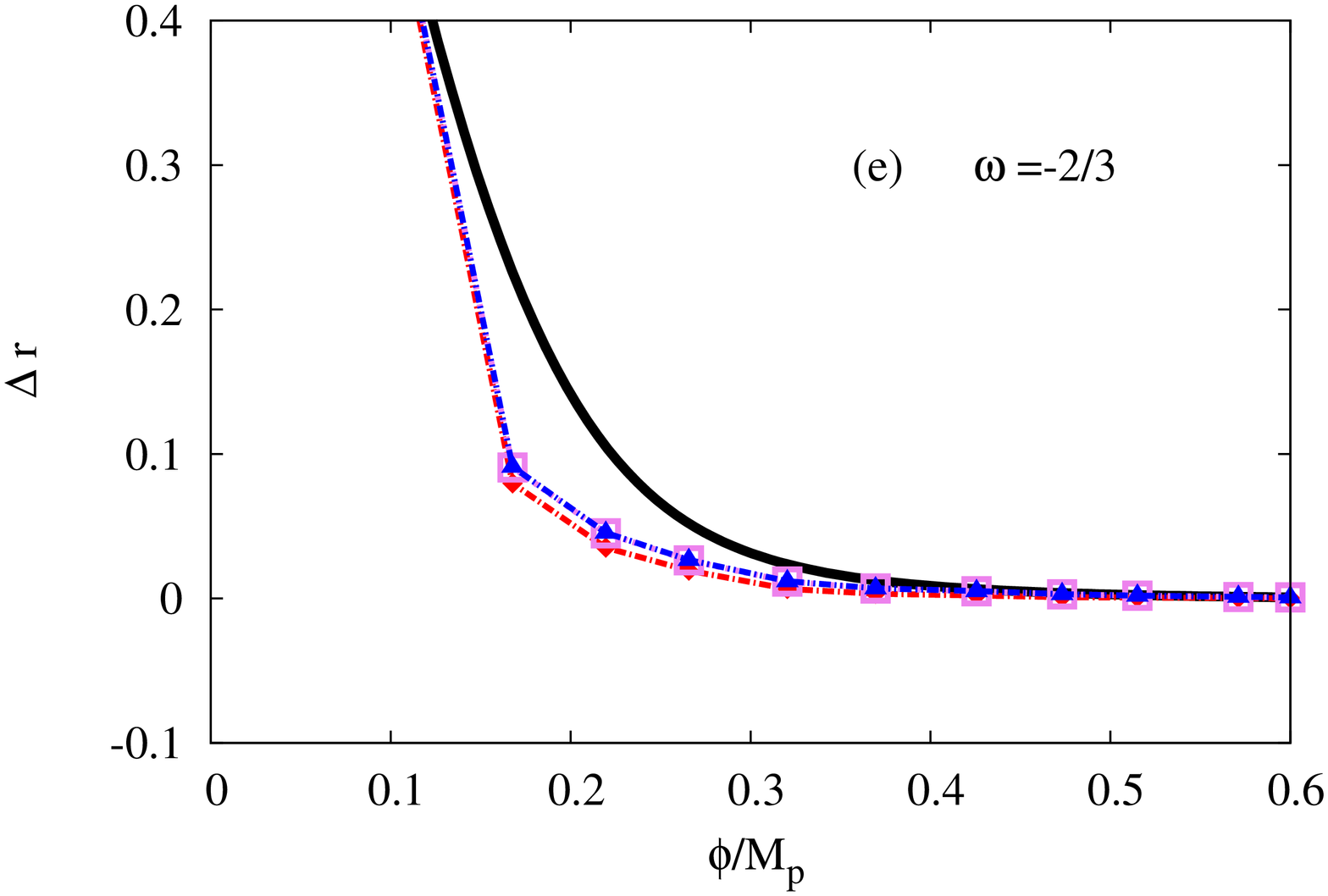}
\includegraphics[width=5.5cm,angle=-90]{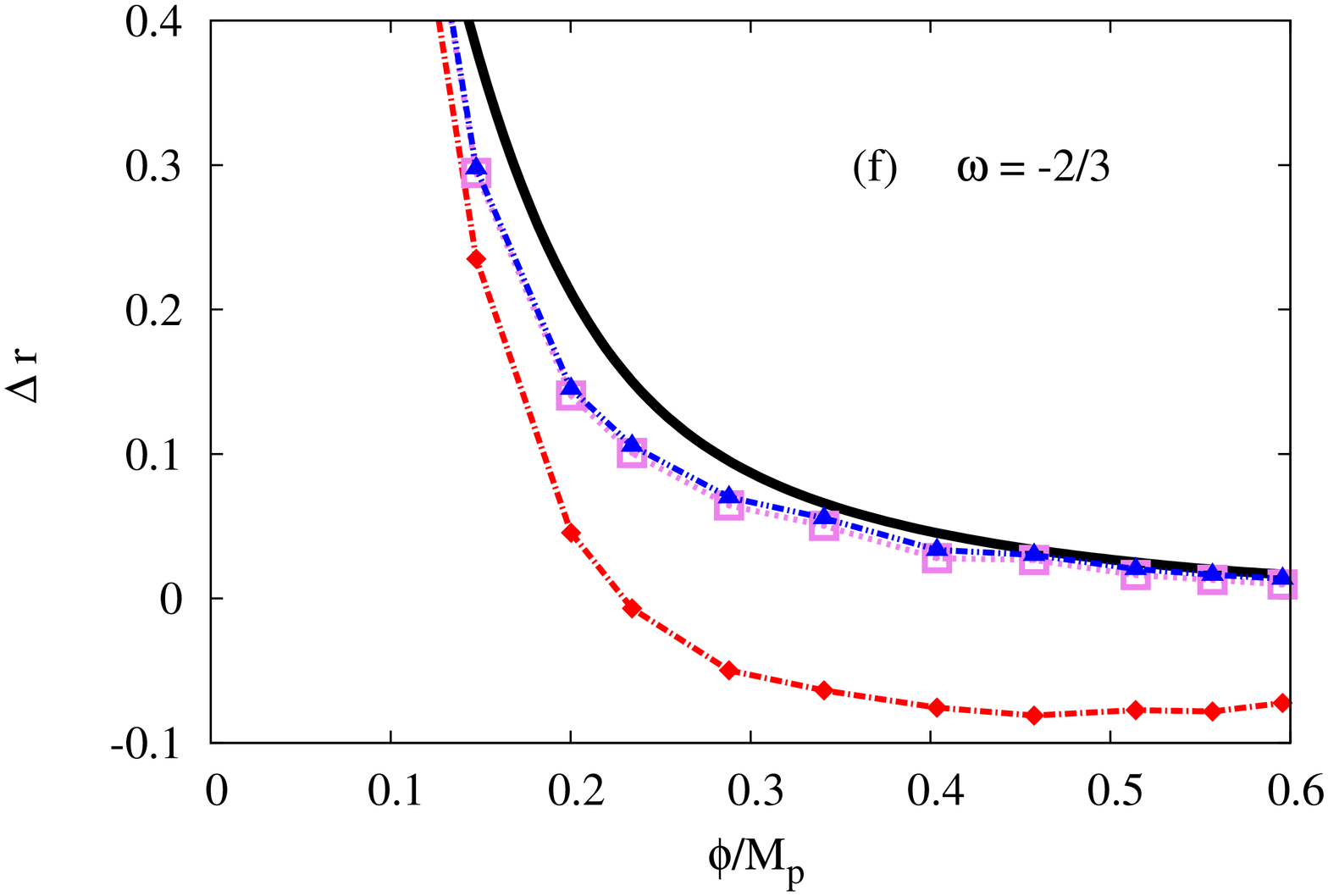}\\
\caption{The same as Fig. \ref{tensorialfluc} but here the net GR-HL impacts [dashed-lines with diamonds (non-detailed), dotted lines with squares (without projectibility) and double-dotted dashed lines with triangles (detailed)] to the inflation parameter $r$, Eq. (\ref{eq:r}), are compared with that from pure GR gravity (solid curve). \label{nsV2} }
}
\end{figure}

The tensorial and scalar density fluctuations, respectively, can be given as \cite{Linde:1982,Liddle:1993,Liddle:1995}
\begin{eqnarray}
P_t &=& \left(\frac{H}{2\pi}\right)^{2} \left[1-\frac{H}{\Lambda}\,\sin\left(\frac{2\Lambda}{H}\right)\right], \label{eq:Pt}\\
P_s &=& \left(\frac{H}{\dot{\phi}}\right)^2\left(\frac{H}{2\pi}\right)^{2} \left[1-\frac{H}{\Lambda}\,\sin\left(\frac{2\Lambda}{H}\right)\right]. \label{eq:Ps} \hspace*{10mm}
\label{ps}
\end{eqnarray} 
Instead, it is more convenient to study the ratio of tensor-to-scalar fluctuations \cite{Linde:1982,Liddle:1993,Liddle:1995}
\bea
r &=& \frac{P_t}{P_s }=\left(\frac{\dot{\phi}}{H}\right)^2, \label{eq:r}
\eea
where the dependence on many parameters can be eliminated. The time derivations of the inflation field are  
\bea
\dot{\phi} &=& \frac{-1}{3H} \partial _{\phi} V(\phi), \\ 
\ddot{\phi} &=&  \frac{- \dot{\phi}}{3 H} \partial ^2 _{\phi} V(\phi). 
\eea

The most interesting aspects of the inflation scenario are the predictions of  the quantum perturbation.  Apart from employing the amplitude of scalar perturbations, we define the scalar spectra index 
\begin{equation}
n_s \equiv 1 - \sqrt{\frac{r}{3}}. \label{eq:ns}
\end{equation}

\begin{figure}
\centering{
\includegraphics[width=5.5cm,angle=-90]{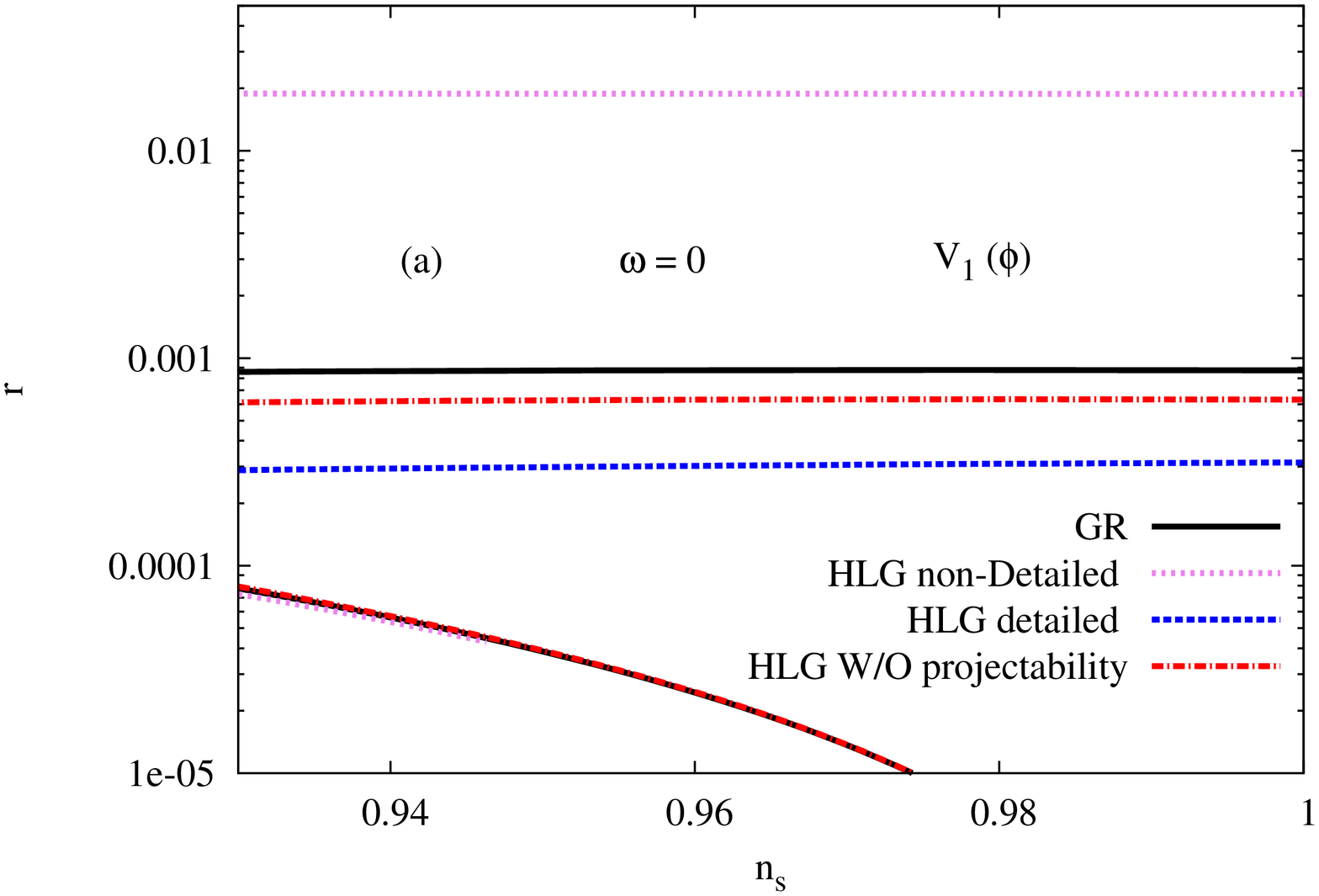}
\includegraphics[width=5.5cm,angle=-90]{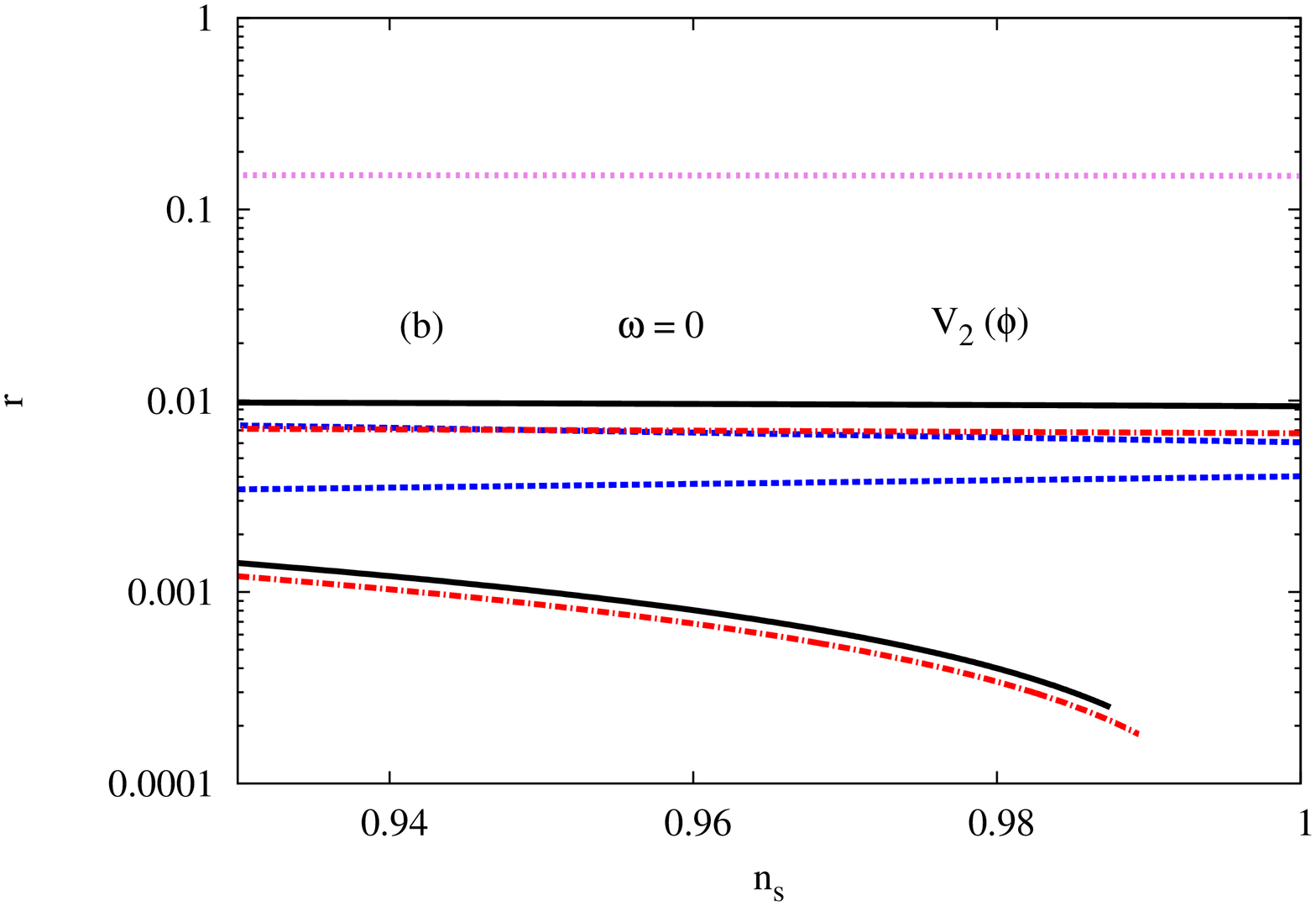}\\
\includegraphics[width=5.5cm,angle=-90]{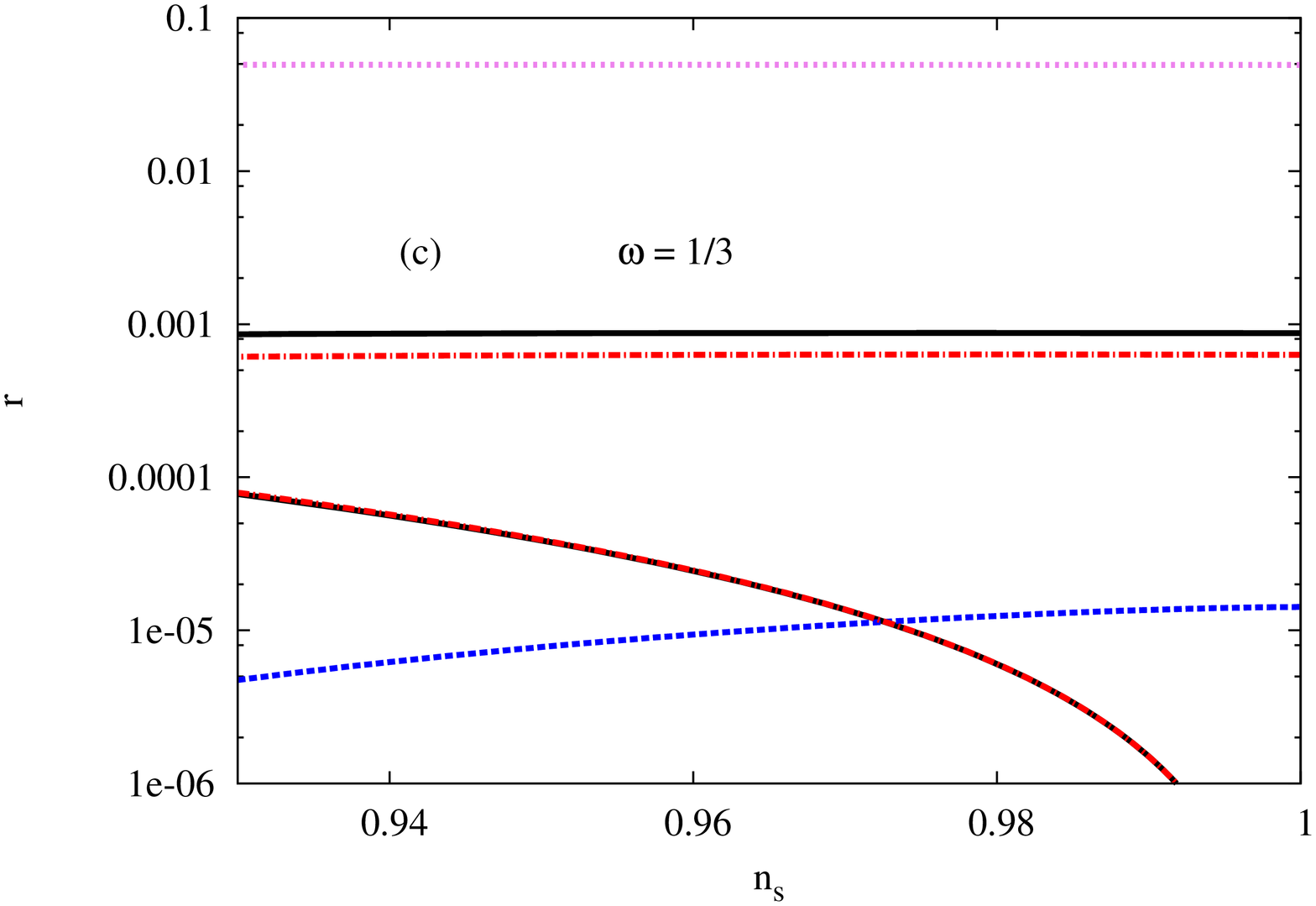}
\includegraphics[width=5.5cm,angle=-90]{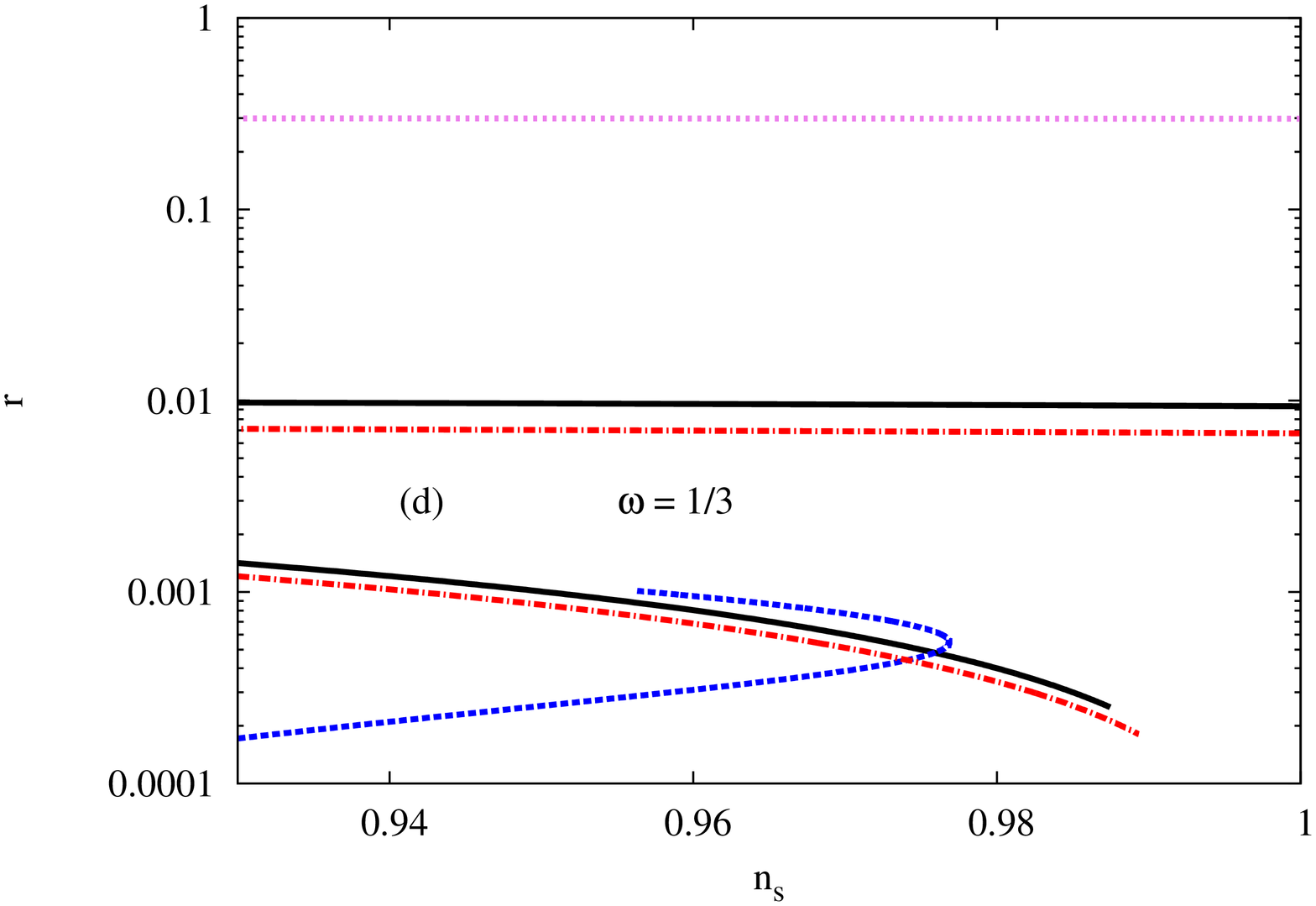}\\
\includegraphics[width=5.5cm,angle=-90]{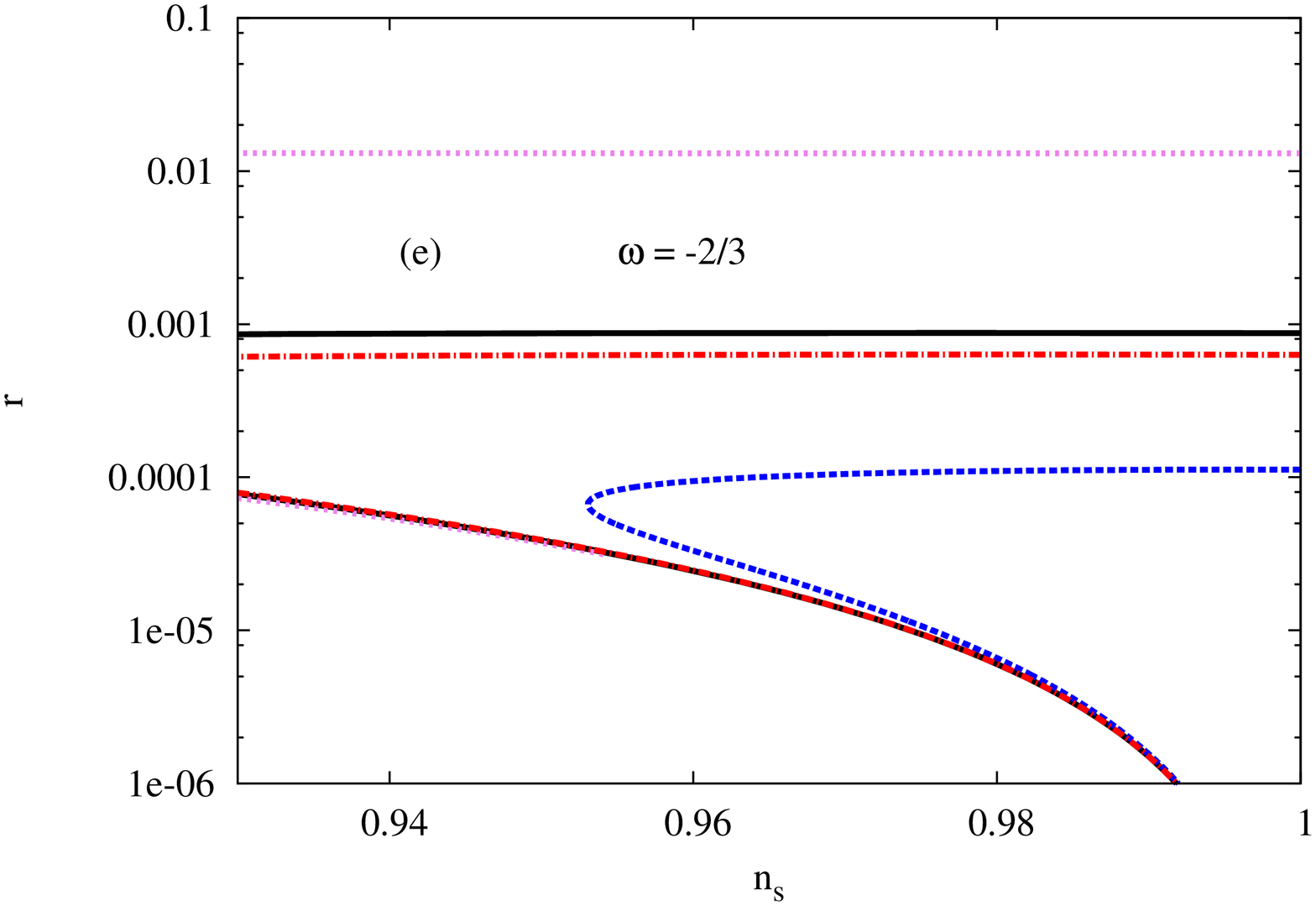}
\includegraphics[width=5.5cm,angle=-90]{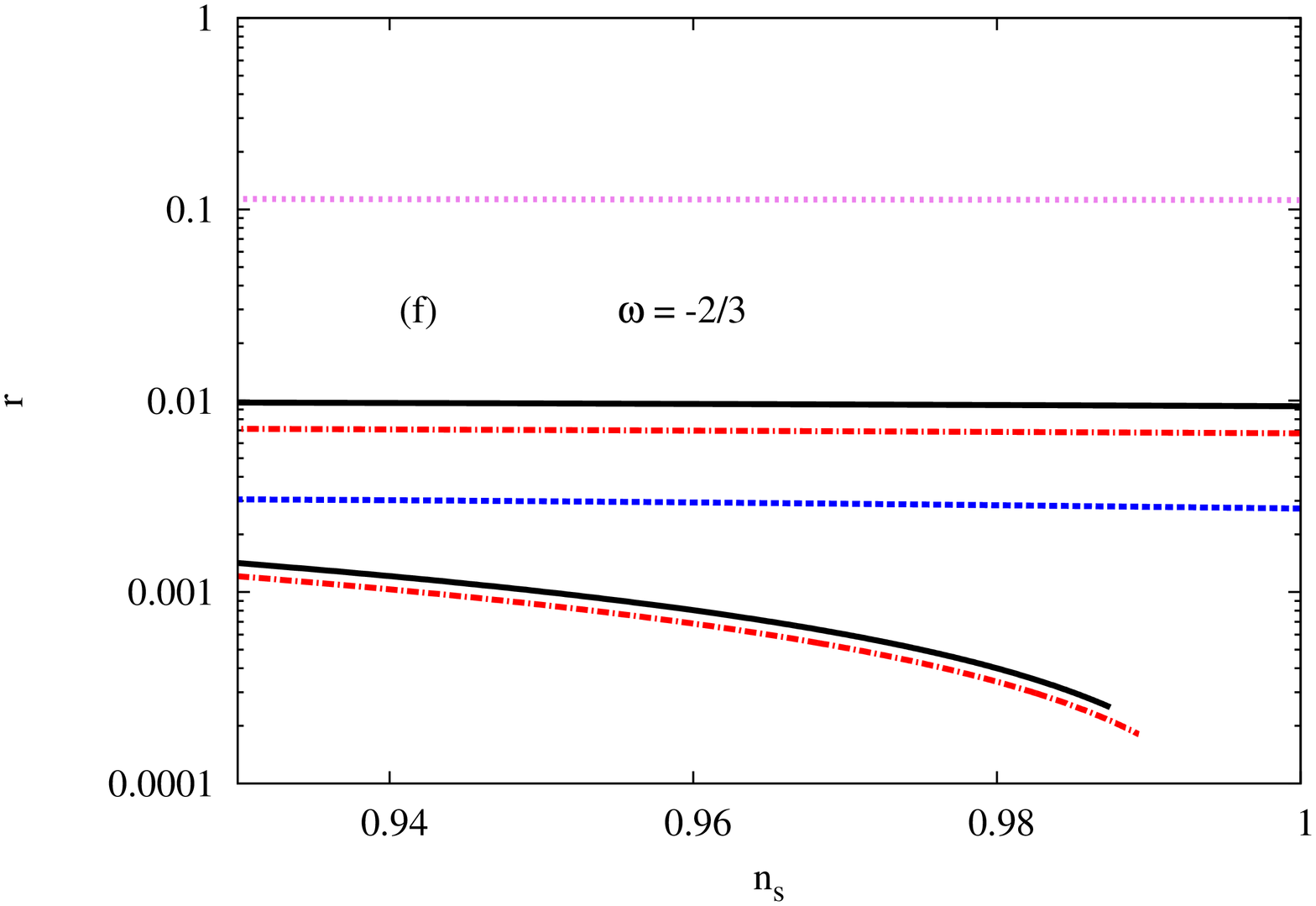}\\
\caption{The same as Fig. \ref{tensorialfluc} but  for tonsorial-to-spectral density fluctuations ($r$) as functions of the spectral index ($n_s$), Eq. (\ref{eq:ns}).}
\label{nsV}
}
\end{figure}

In Fig. \ref{tensorialfluc}, the tonsorial density fluctuations ($P_t$) are given as functions of the inflation field ($\phi$) in units of Planck mass ($M_{p}$), Eq. (\ref{eq:Pt}), for two inflation potentials, Eqs. (\ref{poweri}) and (\ref{eq:mssm}), and different EoS; $\omega =0$ (top panel), $1/3$ (middle panel), and $-2/3$ (bottom panel). The calculations are performed in GR (solid curve), HL gravity detailed (dashed curve), HL gravity non-detailed (dotted curve) and HL gravity without the projectability (dash-dotted curve) condition and implementing two types of inflation potentials; Eq. (\ref{poweri}) in left- and  (\ref{eq:mssm}) in right-hand panel. 

The GR results are almost equivalent to HL gravity without the projectibility condition. Also we notice that $V_1(\phi)$ results in values smaller than that from $V_2(\phi)$. $P_t$ from detailed HL gravity is the smallest, except at $\omega=-2/3$, it becomes the largest. $P_t$ from non-detailed HL gravity with is always smaller that GR (Friedmann gravity).

We find that the dependence of the tonsorial density fluctuations ($P_t$) on the inflation field $\phi$ varies with: 
\begin{itemize}
\item a) the proposed inflation potentials, 
\item b) the equations of state characterizing the cosmic background, and 
\item c) the types/theories of gravity.
\end{itemize}
It is obvious that negative $P_t$ means that the second term in Eq. (\ref{eq:Pt}) should be positive or $\Lambda$ ranges between $\epsilon\, H$ and $(\pi/2)\, H$, where $\epsilon$ is an infinitesimal positive constant. 

Fig. \ref{spectralfluc} shows the same as in Fig. \ref{tensorialfluc}, but for spectral density fluctuations ($P_s$), Eq. (\ref{eq:Ps}). Also here, same conclusions can be drawn. Negative $P_s$ refers to $\epsilon\, H\leq \Lambda \leq (\pi/2)\, H$  and furthermore, $\dot{\phi}\leq 0$ leads to dominant contribution from the cosmological constant. 

Also here, $P_s$ obtained from  GR are almost equivalent to that from HL gravity without the projectibility condition. $P_s$ from detailed HL gravity is the largest, except at $\omega=-2/3$, it becomes the smallest for $V_2(\phi)$ and second largest for $V_1(\phi)$. $P_s$ from non-detailed HL gravity is almost a vanishing constant.   

Fig. \ref{tensorialspectralfluc} presents the ratio of tensorial to spectral density fluctuations ($r=P_t/P_s$), Eq. (\ref{eq:r}) as functions of $n_s$. Again, the inflation fields, the equations of state, and the types of gravity apparently affect the obtained results. It is obvious that eliminating the dependence on various parameters through $r$ seems to bring regularities to all curves. Regularity means concrete patters among both potentials and the equations of state. The largest $r$-results are obtained from non-detailed condition, while detailed condition gives the smallest results. The results from GR and HL gravity without the projectibility condition are similar at least qualitatively, where the earlier lead to larger values than the latter.

In order to determine the role of HL relative to GR gravity, we draw in Fig. \ref{nsV2}  $\Delta r$ which is the differences between HL and GR on $r$ as a function of $\phi/M_p$ from both scalar fields and the three types of the equation of state. The GR-HL results are given as dashed-lines with diamonds (non-detailed HL), dotted lines with squares (HL without projectibility) and double-dotted dashed lines with triangles (detailed HL)]. For a better comparison, we also illustrate the pure GR contributions as solid curves.  

When $V_1(\phi)$ is taken into consideration, it is obvious that $\Delta r$ is almost identical for all equations of state. Furthermore, their values are very close to that of pure GR gravity. This means that HL gravity comes up with negligibly small contributions, especially at $\omega=0$. At $\omega=1/3$ and $-2/3$, we find that HL results on $\Delta r$ become faster than that of the pure GR gravity, especially when the scalar fields range between $10\%$ to $\sim 25\%$ of $M_p$.   

For $V_2(\phi)$, there is a noticeable differ between $\Delta r$ from non-detailed HL and pure GR gravity. The rate of  $\Delta r$ decreasing with increasing $\phi/M_p$ becomes faster relative the all other results.  At large $\phi/M_p$, the contributions from the non-detailed HL gravity become dominant (negative $\Delta r$). 

When the cosmic background geometry is characterized by $\omega=0$, HL has a tiny impact, except for non-detailed condition in $V_2(\phi)$. Apart from the remarkable impacts of non-detailed HL gravity, the differences between GR and other HL theories become noticeable at $0.1<\phi< 0.25\, M_p$ for $\omega=1/3$ and $-2/3$.

\subsection{Recent PLANCK Observations}
\label{sec:planck}

\begin{figure}
\begin{center}
\includegraphics[width=12.cm,angle=-0]{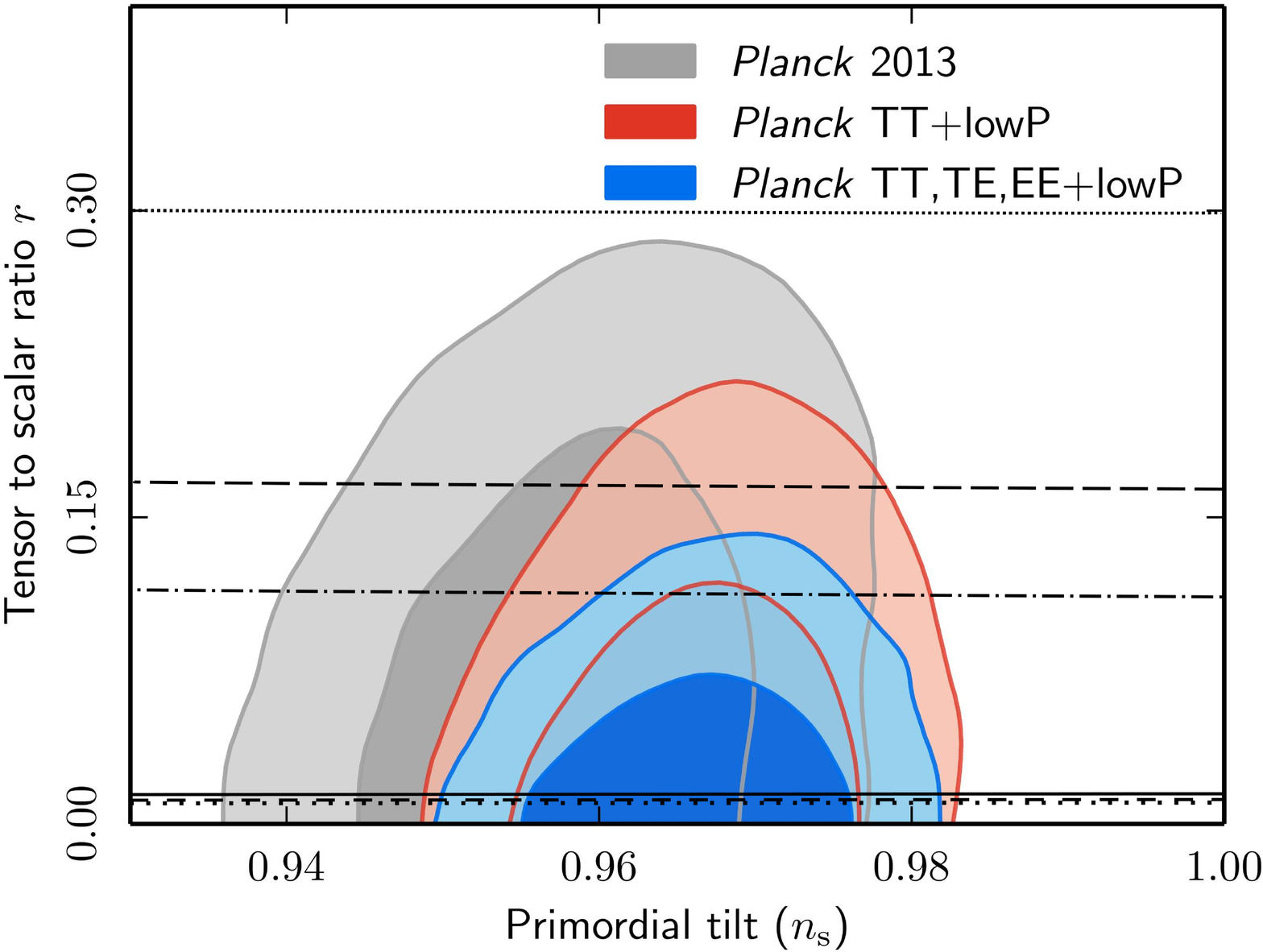} 
\caption{Some results illustrated in Fig. \ref{nsV} are confronted here to the recent PLANCK observations (red and blue contours), which is also compared to previous observations (gray contours). The horizontal lines represents largest $r$ obtained through parametric calculations from $V_1(\phi)$ (bottom three lines) with non-detailed HL gravity at $\omega=1/3$ (solid), $\omega=0$ (dashed) and $\omega=-2/3$ (dotted line). The results  from $V_2(\phi)$ (top three lines) with non-detailed HL gravity at $\omega=1/3$ (double dotted), $\omega=0$ (long dashed) and $\omega=-2/3$ (dot-dashed line).  }
\label{rVsns}
\end{center}
\end{figure}

In Fig. \ref{nsV}, the tonsorial-to-spectral density fluctuations ratio ($r$) are depicted as functions of the spectral index ($n_s$). Recently, such a dependence has been analysed by PLANCK collaboration, Fig. \ref{rVsns}, as well. 

In Fig. \ref{rVsns}, some of the results deduced from the parametric dependence of $r$ on the spectral index ($n_s$) are confronted to the recent PLANCK. Obviously, they fit well with the observations at $68\%$ and $95\%$ confidence level \cite{planck2015a,planck2015b}. The figure also illustrates a significant improvement (red and blue contours) with respect to previous PLANCK data release (gray contours). 

The recent PLANCK observations set an upper bound to the tensor-to-scalar ratio, $r_{0.002}<0.11$  at $95\%$ confidence level, when taking into consideration the PLANCK high-$\ell$ polarization data. The upper limit according to {\it B}-mode polarization constraint, $r< 0.12$ at $95\,\%$ at confidence level, which was obtained from a joint analysis of PLANCK, BICEP2, and Keck Array data \cite{planck2015a} is well reproduced. 

In Fig. \ref{rVsns}, we draw the largest $r$ as obtained from our parametric calculations (solid line). This was deduced from non-detailed HL gravity.  As shown in Fig. \ref{nsV2}, only non-detailed HL gravity causes a noticeable impact relative to the pure GR gravity.  We depict largest $r$ calculated from $V_1(\phi)$ as the bottom three lines; at $\omega=1/3$ (solid), $\omega=0$ (dashed) and $\omega=-2/3$ (dotted line).  The small values can be interpreted due to the negligibly small $\Delta r$, Fig. \ref{nsV2}. The results from $V_2(\phi)$ are presented by the top three lines; at $\omega=1/3$ (double dotted), $\omega=0$ (long dashed) and $\omega=-2/3$ (dot-dashed line. It is obvious that almost all our calculations fit well with the recent observations, except from $V_2(\phi)$ at $\omega=1/3$ (double dotted). Remaining results on $r$ vs. $n_s$ agree well with the recent observations. Among other theories of gravity, it is obvious that non-detailed HL gravity results in the largest $r$, e.g. $1-2$ orders of magnitudes.

\section{Conclusions}
\label{sec:conc}

Based on the standard model of elementary particles, i.e. the successful description of all forces (except gravity) by quantum field theory, and Lifshitz theory with an anisotropic scaling in UV between Minkowskian space and time which was based on a scalar field theory for condensed matter physics, Horava proposed a theory for quantum gravity with an anisotropic scaling in UV. He even concluded that such modification in gravity is responsible for the accelerated expansion of the Universe. In GR and various approaches for HL gravity, we have studied the impacts of different equations-of-state on Friedmann inflation with a scalar field. HL gravity possesses power-counting renormalizablity and assures causal dynamical triangulations. Its renormalization is based on asymptotic safety and symmetries of GR. HL gravity introduces a new set of symmetries imposing invariance under foliation-preserving diffeomorphisms. 

The present work describes a systematic analysis for the effects of different equations-of-state characterizing the cosmic background geometry and compares between the conventional gravity characterized by GR and its HL counterparts; detailed, and non-detailed HL gravity, and that without the projectability condition. 

We have studied the dependence of tensorial and spectral density fluctuations on two scaler fields, $V_1(\phi)$ [Eq. (\ref{poweri}) and $V_2(\phi)$ Eq. (\ref{eq:mssm})]. The latter are imposed to assure inflationary mechanism. It was found that the proposed inflation potentials, and/or the equations of state characterizing the cosmic background geometry, and/or gravity theories control how density fluctuations vary with the scaler field. We found that the GR results on tensorial ($P_t$) and spectral density fluctuations ($P_s$) are almost equivalent to HL gravity without the projectibility condition, at least qualitatively. Also we notice that $V_1(\phi)$ results in values smaller than the ones deduced from $V_2(\phi)$. The results on both $P_t$ and  $P_s$ from detailed and non-detailed HL gravity strongly depend on the equation of state and of course on the inflation potential field.

The dependence of the tensorial-to-spectral density fluctuations ($r$) on the quantum perturbations in form of scalar spectra index has been calculated. There is a regular pattern characterizing both potentials and the equations of state. The largest results are obtained from non-detailed HL gravity, while detailed condition gives the smallest results. The results from GR and HL gravity without the projectibility condition are almost identical. The earlier lead to larger values than the latter.

We conclude that negative tensorial fluctuations refer to dominant contributions by the cosmological constant. Negative spectral fluctuations summarize that effect plus that from scalar field temporal evolution. Also, we find that the tensorial-to-spectral density fluctuations contentiously decrease when moving from non-detailed HL gravity, to GR gravity, to HL gravity without the projectibility condition, and to detailed HL gravity. Last but not least, we find that GR results are very close to that from HL gravity without the projectibility condition.

Our calculations on the tensorial-to-spectral density fluctuations are compared with the recent PLANCK observations, i.e. $r_{0.002}<0.11$  at $95\%$ confidence level in PLANCK high-$\ell$ polarization data. The upper limit according to {\it B}-mode polarization constrains $r$ to $< 0.12$ at $95\,\%$ at confidence level. The latter was obtained from a joint analysis of PLANCK, BICEP2, and Keck Array data \cite{planck2015a}.  It is obvious that our parametric calculations agree well with the recent PLANCK observations.

At $\omega=0$, HL gravities have tiny impacts, except HL with non-detailed condition in scalar field $V_2(\phi)$. Apart from the dominant role of non-detailed HL relative to GR gravity, the differences between GR and other HL theories are only distinguishable from the pure GR at $0.1<\phi< 0.25\, M_p$ for $\omega=1/3$ and $-2/3$.

Last but not least,  it seems that introducing {\it extra} degrees of freedom in forms of additional scalar fields which were assumed in order to derive the inflation, enables us to describe well the recent PLACK observations.


\begin{thebibliography}{99}

\bibitem[\protect\citeauthoryear{Ade}{2015a}]{planck2015a} P.A.R. Ade {\it et al.} (BICEP2/Keck and Planck Collaborations), Phys. Rev. Lett. {\bf 114}, 101301 (2015).
\bibitem[\protect\citeauthoryear{Ade}{2015b}]{planck2015b}  P.A.R. Ade {\it et al.} (Planck Collaboration), {\it ''Planck 2015 results XX. Constraints on inflation''}, 1502.02114 [astro-ph.CO].
\bibitem[\protect\citeauthoryear{Albrecht}{1982}]{infl2} A. Albrecht, and P. J. Steinhardt, Phys. Rev. Lett. {\bf 48}, 1220 (1982).
\bibitem[\protect\citeauthoryear{Allahverdi}{2006}]{allahverdi-2006} R. Allahverdi, J. Garcia-Bellido, K. Enqvist, and A. Mazumdar, Phys. Rev. Lett. {\bf 97}, 191304 (2006).

\bibitem[\protect\citeauthoryear{Barrow}{1981}]{infl}  J.D. Barrow, and M.S. Turner, 
Nature {\bf 292}, 35 (1981).
\bibitem[\protect\citeauthoryear{Blas}{2010}]{const1a} D. Blas, O. Pujolas, and S. Sibiryakov, Phys. Rev. Lett. {\bf 104}, 181302 (2010).  
\bibitem[\protect\citeauthoryear{Blas}{2011}]{const1b} D. Blas, O. Pujolas, and S. Sibiryakov, J. High Energy Phys, {\bf 1104}, 018 (2011). 


\bibitem{Abdalla:2005} M.C.B. Abdalla, S. Nojiri, S. D. Odintsov, Class. Quant. Grav. {\bf 22}, L35 (2005). 

\bibitem{Sasaki:2010a} Jinn-Ouk Gong, S. Koh, M. Sasaki,  Phys. Rev. D {\bf 81}, 084053 (2010).

\bibitem{Calcagni:2009s} G. Calcagni, JHEP {\bf 0909}, 112 (2009).

\bibitem{Kiritsis:2009a} E. Kiritsis, G. Kofinas, Nucl. Phys. B {\bf 821}, 467 (2009).



\bibitem[\protect\citeauthoryear{Calcagni}{2009}]{Reff7} G. Calcagni, J. High Energy Phys. {\bf 09}, 112 (2009). 
\bibitem[\protect\citeauthoryear{Chaichain}{2012}]{masud} M. Chaichian, Sh. Nojiri, S. D. Odintsov, M. Oksanen, and A. Tureanu,  Class. Quant. Grav. {\bf 27}, 185021 (2010), Erratum-ibid. {\bf 29}, 159501 (2012). 
\bibitem[\protect\citeauthoryear{Charmousis}{2009}]{Charm} C. Charmousis, G. Niz, A. Padilla and P.M. Saffin, 
JHEP {\bf 0908}, 070 (2009). 
\bibitem[\protect\citeauthoryear{Chibisov}{1982}]{qfluct1} G. Chibisov and V. Mukhanov, 
 Mon. Not. R. Astron. Soc. {\bf 200} 535 (1982).

\bibitem[\protect\citeauthoryear{Elizalde}{2010}]{1006.3387}  
E. Elizalde, S. Nojiri, S.D. Odintsov, and D. Saez-Gomez, 
Eur. Phys. J. C {\bf 70}, 351 (2010). 

\bibitem[\protect\citeauthoryear{Guth}{1981}]{infl1}  A.H. Guth, Phys. Rev. D {\bf 23}, 347 (1981).
\bibitem[\protect\citeauthoryear{Guth}{1998}]{inflaton} A.H. Guth, {\it ''The Inflationary Universe: The Quest for a New Theory of Cosmic Origins''}, (Vintage, London, 1998).

\bibitem[\protect\citeauthoryear{Horava}{2009}]{Horava2009a} P. Horava, 
JHEP {\bf 0903}, 020 (2009). 
\bibitem[\protect\citeauthoryear{Horava}{2011}]{Horava2011a} P. Horava, 
Class. Quant. Grav. {\bf 28}, 114012 (2011). 
\bibitem[\protect\citeauthoryear{Huang}{2012}]{1201.4630}  
Y. Huang, Anzhong Wang, and Qiang Wu, 
JCAP {\bf 1210}, 010 (2012). 

\bibitem[\protect\citeauthoryear{Kase}{2015}]{1409.1984} R. Kase, and S. Tsujikawa, 
Int. J. Mod. Phys. D {\bf 23}, 1443008  (2015). 
\bibitem[\protect\citeauthoryear{Kazanas}{1980}]{inflhp}  L. Kazanas, 
Astrophys. J. {\bf 241}, L59 (1980).
\bibitem[\protect\citeauthoryear{Kheyri}{2013}]{ir1} F. Kheyri, M. Khodadi, and H.R. Sepangi, 
Eur. Phys. J. C {\bf 73}, 2286 (2013).
\bibitem[\protect\citeauthoryear{Kimpton}{2010}]{const2} I. Kimpton and A. Padilla,  J. High Energy Phys. {\bf 07}, 014 (2010). 

\bibitem[\protect\citeauthoryear{Liddle}{1993}]{Liddle:1993} A.R. Liddle, and D. H. Lyth, Phys. Rept. {\bf 231}, 1 (1993).
\bibitem[\protect\citeauthoryear{Liddle}{2003}]{Liddle:2003} A.R. Liddle, {\it ''Introduction to modern cosmology''} (Wiley, UK, 2003).
\bibitem[\protect\citeauthoryear{Lidsey}{1997}]{Liddle:1995} J.E. Lidsey, A. R. Liddle, E. W. Kolb, E. J. Copeland, T. Barreiro, and M. Abney, Rev. Mod. Phys. {\bf 69}, 373 (1997).
\bibitem[\protect\citeauthoryear{Lifshitz}{1941}]{lifshitz} E.M. Lifshitz, ZH. Eksp. Toer. Fiz. {\bf 11}, 255 (1941).
\bibitem[\protect\citeauthoryear{Linde}{1982}]{Linde:1982} A.D. Linde, Phys. Lett. B {\bf 108}, 389 (1982).
\bibitem[\protect\citeauthoryear{Linde}{1983}]{monop} A.D. Linde, 
Phys. Lett. B {\bf 132}, 317 (1983).
\bibitem[\protect\citeauthoryear{Linde}{2002}]{Linde:2002} A.D. Linde, (ed. S. Bonometto, V. Gorini and U. Moschella) {\it ''Inflationary cosmology and creation of matter in the Universe. In Modern cosmology''}, (Institute of Physics Publishing, Bristol, 2002).
\bibitem[\protect\citeauthoryear{Lukash}{1980}]{qfluct2}  V. Lukash, Pis’ma Zh. Eksp. Teor. Fiz. {\bf 31}, 631 (1980).
\bibitem[\protect\citeauthoryear{L\"u}{2009}]{Reff6}  H. L\"u, J. Mei, and C.N. Pope, Phys. Rev. Lett. {\bf 103}, 091301 (2009). 

\bibitem[\protect\citeauthoryear{Nojiri}{2006}]{nori} S. Nojiri and S. D. Odintsov, Int. J. Geom. Meth. Mod. Phys. {\bf 4}, 115 (2007). 

\bibitem[\protect\citeauthoryear{Mukohyama}{2009}]{0904.2190} S. Mukohyama, 
JCAP {\bf 0906}, 001 (2009). 

\bibitem[\protect\citeauthoryear{Press}{1980}]{inflnt1} W. Press, Phys. Scr. {\bf 21}, 702 (1980).

\bibitem[\protect\citeauthoryear{Sanchez}{2007}]{sanchez-2007} J.C.B. Sanchez, K. Dimopoulos, and D. H. Lyth, JCAP {\bf 0701}, 015 (2007).
\bibitem[\protect\citeauthoryear{Sato}{1981}]{inflnt2} K. Sato, Mon. Not. R. Astron. Soc. {\bf 195}, 467 (1981).
\bibitem[\protect\citeauthoryear{Starobinsky}{1980}]{Starobinsky}  A. A. Starobinsky,  
Phys. Lett. B {\bf 91},  99 (1980).
\bibitem[\protect\citeauthoryear{Steinhardt}{2011}]{inflobs1} P. J. Steinhardt, 
Scientific American {\bf 4}, 18 (2011). 

\bibitem[\protect\citeauthoryear{Tawfik}{2015}]{TDE} A. Tawfik and E. Abou El Dahab, {\it ''Friedmann-Lemaitre-Robertson-Walker Cosmology with Horava-Lifshitz Gravity: Impacts of various Equations of State''}, submitted to INJP

\bibitem[\protect\citeauthoryear{Tsujikawa}{2003}]{inflobs2} S. Tsujikawa, {\it ''Introductory review of cosmic inflation''},  hep-ph/0304257. 

\bibitem[\protect\citeauthoryear{Verlinde}{2011}]{Verlinde2011a} E.P. Verlinde, 
J. High Energy Phys. {\bf 1104}, 029 (2011). 

\bibitem[\protect\citeauthoryear{Zhu}{2011}]{modl1} T. Zhu, Q. Wu, A. Wang and F.-W. Shu, 
Phys. Rev. D {\bf 84}, 101502 (2011). 
\bibitem[\protect\citeauthoryear{Zhu}{2012}]{modl2} T. Zhu, F.-W. Shu, Q. Wu and A. Wang, 
Phys. Rev. D {\bf 85}, 044053 (2012). 
\bibitem[\protect\citeauthoryear{Zhu}{2013}]{1208.2491}  
T. Zhu, Yongqing Huang, and Anzhong Wang, 
JHEP {\bf 1301}, 138 (2013). 






























  

  
  
\end{thebibliography}
\end{document}